\documentclass[%
 reprint,
 amsmath,amssymb,
 aps,
]{revtex4-2}

\usepackage{graphicx}
\usepackage{dcolumn}
\usepackage{bm}
\usepackage{hyperref}
\usepackage{xcolor}
\usepackage{float}
\usepackage{tikz}
\DeclareMathOperator\arctanh{arctanh}

\begin{document}

\preprint{APS/123-QED}

\title{Equilibrium cluster statistics of cooperative and anticooperative binding on finite one-dimensional rings}
\author{Thomas Alfonsi}
\email{thomas.alfonsi@umontpellier.fr}
\affiliation{
Laboratoire Charles Coulomb (L2C), Univ. Montpellier, CNRS UMR5221, Montpellier, France
}
\author{Jérôme Dorignac}
\affiliation{
Laboratoire Charles Coulomb (L2C), Univ. Montpellier, CNRS UMR5221, Montpellier, France
}
\author{John Palmeri}
\affiliation{
Laboratoire Charles Coulomb (L2C), Univ. Montpellier, CNRS UMR5221, Montpellier, France
}
\author{Nils-Ole Walliser}
\email{nils-ole.walliser@umontpellier.fr}
\affiliation{
Laboratoire Charles Coulomb (L2C), Univ. Montpellier, CNRS UMR5221, Montpellier, France
}

\date{\today}
\begin{abstract}
We study equilibrium clustering in a finite one-dimensional lattice gas of $L$ sites with periodic boundary conditions, as a minimal model for adsorption and binding on small ring-like substrates. 
Using a grand-canonical formulation with nearest-neighbor coupling, we derive exact finite-size expressions for the mean occupancy, the mean number of domain walls, and the mean number of clusters. 
Building on exact $k$-site correlation functions, we further derive expressions for the mean number of clusters of size $k$ and for two complementary size statistics: the cluster-size distribution, and the site-weighted cluster-size distribution.
These observables characterize how spatial organization changes across attractive (cooperative) and repulsive (anticooperative) interactions, and highlight finite-size and parity-dependent effects of the underlying lattice, the latter being particularly pronounced near half filling in small systems.
To access larger lattices without enumerating all $2^L$ microstates, we also develop a cluster-based combinatorial formulation in which configurations are classified by cluster counts and sizes, reducing the effective state space to a set whose size scales with integer partitions, $\approx e^{\sqrt{L}}$, rather than with $\approx e^{L}$.
Taken together, our results provide exact benchmarks for finite periodic systems and suggest experimentally relevant cluster observables that complement occupancy-based measures of cooperativity, with particular relevance for binding on ring-like substrates for biological assemblies.

\end{abstract}

\maketitle

\section{\label{sec:intro}Introduction}
The present work is motivated by the need to understand the spatial organization of bound units in adsorption processes on small one-dimensional periodic systems.
A cluster-based description is particularly well-suited for this purpose within the short-range lattice gas (SRLG) model, where only nearest-neighbor interactions are considered. 

The formation of clusters has been extensively studied for two- and three-dimensional Ising systems \cite{binder1976clusters}.
\textcite{liu1993application} used the Ising model to study the cooperativity between ion channels and analyzed the the mean relative occupancy and its fluctuations for a one-dimensional lattice at equilibrium.
More generally, the spatial organization of the particles on a one-dimensional lattice has been widely investigated within both Ising and lattice-gas frameworks.
These studies led to important results such as explicit analytical expressions for the number of clusters of a given size in the thermodynamic limit \cite{yilmaz2005exact}; approximated forms of the probability distribution function of clusters at low densities on infinite lattices \cite{fronczak2013cluster}; exact results for open lattices both at finite size and in the thermodynamic limit \cite{ivanytskyi2018bimodal}; and exact results for finite periodic lattices in the canonical ensemble \cite{vavro2001exact}.

We expand on these previous results and propose a unified analysis of equilibrium clustering in a one-dimensional SRLG on a periodic lattice of size $L$ within the grand-canonical ensemble. We derive explicit relations at finite $L$ for a broad set of cluster observables, including the mean number of domain walls and particle clusters, cluster-size statistics (site-weighted and cluster-weighted), and mean cluster-size measures, and we quantify finite-size and parity effects that persist near half filling. We connect these results to instructive limits, including half filling and noninteracting (Hill-Langmuir) case,  and we clarify the physical interpretation of the different parameter regimes and the crossover between strongly attractive (cooperative) and strongly repulsive (anticooperative) interactions.

Our focus on moderate lattice sizes ($L\sim 10$), far from the thermodynamic limit, is motivated by their increasing experimental relevance, in particular in biological settings where recent advances may allow cluster statistics to be probed \cite{reid2006maximum, nord2017catch}. Technically, we obtain the finite-size results using two complementary approaches. First, a site-based transfer-matrix/correlation-function formalism yields closed expressions at fixed $L$ (Sec.~\ref{sec: occupation formalism}). Second, a cluster-based combinatorial formalism enumerates configurations via cluster-size multiplicities, reducing the effective state space to a size scaling with the integer partition number $p(L)$ rather than with the $2^L$ microscopic site configurations (Sec.~\ref{sec:cluster formalism}), thereby enabling exact calculations for larger systems.

Beyond providing exact expressions, our results yield exact mappings between equilibrium observables, producing parametric ``fingerprints'' that relate the mean occupancy to cluster descriptors (e.g., mean cluster number and mean cluster sizes) in finite periodic systems. Because the number of clusters and domain walls depends only on the cluster decomposition, and does not require single-particle resolution within clusters, combining measurements of $\langle\varphi\rangle$ with a cluster observable can, in principle, constrain the range of compatible system parameters (effective interaction strength and chemical potential), thereby complementing inference approaches based solely on occupancy fluctuations \cite{franco2025signature}.
\section{\label{sec: occupation formalism}Occupation formalism}
Let us consider a 1D lattice with periodic boundary conditions and $L$ lattice sites in contact with a thermal bath at temperature $T$ and a particle reservoir with a constant chemical potential $\mu_r$.
The Hamiltonian of the Short Range Lattice Gas (SRLG) can be written as \cite{friedli_velenik_2017,franco2025signature}
\begin{equation}
    \beta \mathcal{H}(\bm{\varphi}) = -J \sum^{L}_{i=1} \varphi_i \varphi_{i+1} - \mu \sum^{L}_{i=1} \varphi_i\, ,
\label{eq: hamiltonian sites}
\end{equation}
where $\varphi_i=0$ if site $i$ is empty and $\varphi_i=1$ if it is occupied. Note that, due to periodic boundary conditions, $\varphi_{L+1} \equiv \varphi_1$.
The parameter $J$ is the dimensionless nearest neighbor interaction potential, which, from now on, we will employ as a measure of the system's cooperative ($J>0$) or anticooperative ($J<0$) character, and $\mu$ is the dimensionless effective chemical potential of the system, defined as $\mu = \mu_r - \epsilon$, where $\epsilon$ is the binding energy ($\epsilon < 0$).
$\beta = 1/k_{\text{B}}T$ with $k_{\text{B}}$, the Boltzmann constant.
The array $\bm{\varphi} = \{\varphi_i\} = \{\varphi_1, ...,\varphi_L \}$ denotes a unique configuration of the lattice out of the $2^L$ possible microscopic ones. An important quantity, the relative occupancy of a configuration, is defined by $\varphi = L^{-1} \sum^L_{i=1} \varphi_i$. The mesostate of this system is defined by the number of particles attached to the lattice $N = L\varphi$. The number of configurations for each mesostate can be computed via the binomial coefficient $ L! /[N! (L-N)!]$ which gives the multiplicity of each mesostate.
The grand canonical partition function for the SRLG can be written as \cite{kramers1941statistics,kardar2007statistical}
\begin{equation}
  \Xi := \sum_{\bm{\varphi}} e^{-\beta \mathcal{H}(\bm{\varphi}) } = \sum_{\bm{\varphi}} \prod_{i=1}^L \langle \varphi_i | T | \varphi_{i+1} \rangle \, ,
\label{eq: partition function}
\end{equation}
where the transfer matrix $T$ is given by
\begin{equation}
  T = 
  \begin{pmatrix}
  1 & e^{\mu/2} \\
  e^{\mu/2} & e^{(J+\mu)}
  \end{pmatrix}\, ,
\label{eq: transfer matrix}
\end{equation} 
and the kets denoting an empty or occupied site are given by $| \varphi = 0 \rangle = \binom{1}{0}$ and $|\varphi = 1 \rangle = \binom{0}{1}$, respectively.
With periodic boundary conditions, the grand canonical partition function can be written as $\Xi = \text{Tr } \left(T^L\right) = \lambda_+^L + \lambda_-^L$, where $\lambda_{\pm}$ are the eigenvalues of the transfer matrix given by
\begin{equation}
    \lambda_{\pm} = e^X \left(\cosh X \pm \sqrt{\sinh^2 X + e^{-J}}\right),
    \label{eq: eigenvalues}
\end{equation}
and where we have defined the variable $X$ as 
\begin{equation}
    X = \frac{J+\mu}{2}.
\end{equation}
\subsection{\label{subsec: mean relative occupancy}Mean relative occupancy}
At equilibrium, the mean relative occupancy of the system is given by 
\begin{align}
  \langle \varphi \rangle &= (L\Xi)^{-1}\partial_\mu \Xi \nonumber \\
  &=\frac{1}{2} \left( 1+\tanh\left({\frac{L}{2\xi}}\right) \frac{\sinh{X}}{\sqrt{\sinh^2{X} + e^{-J}}} \right)\, ,
\label{eq: mean relative occupancy}
\end{align}
where $\xi$ is the correlation length of the system defined by
\begin{equation}
  \xi = 1/\log\left(\lambda_+/\lambda_-\right) \, .
\label{correlation length}
\end{equation}
With this definition, we can also define the mean number of bound particles at equilibrium as $\langle N \rangle = L \langle \varphi \rangle$.
The mean relative occupancy will be used as a control parameter in the system, giving access to $\mu$ when $J$ is fixed.
All results shown correspond to equilibrium conditions unless explicitly stated otherwise; accordingly, observables written with angle brackets represent equilibrium averages.
\begin{figure}[h]
\centering
\begin{tikzpicture}
\node (img) at (0,0)
    {\includegraphics[width=0.5\textwidth]{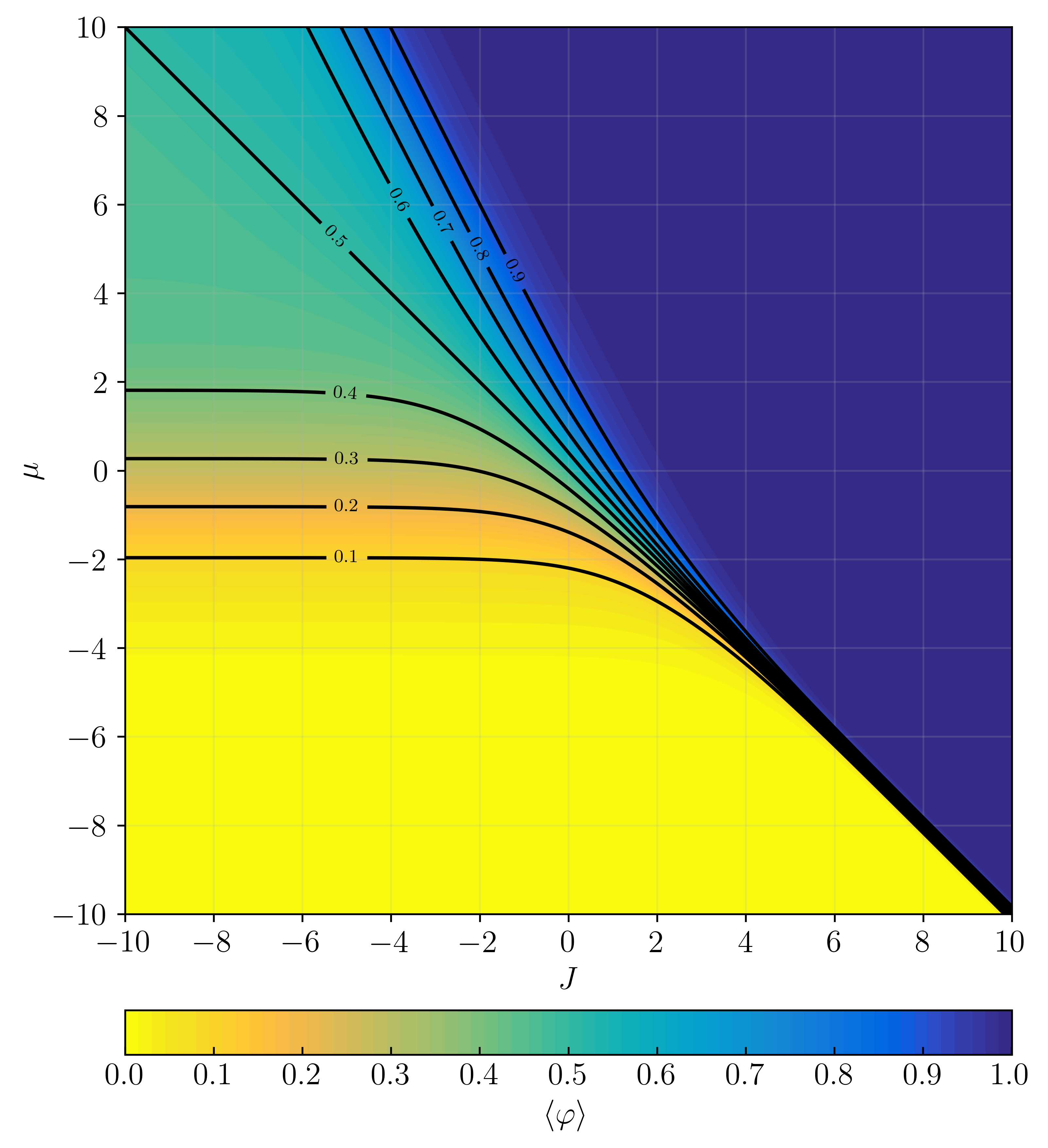}};
\node at (2.4,2.9)
    {\includegraphics[width=0.18\textwidth]{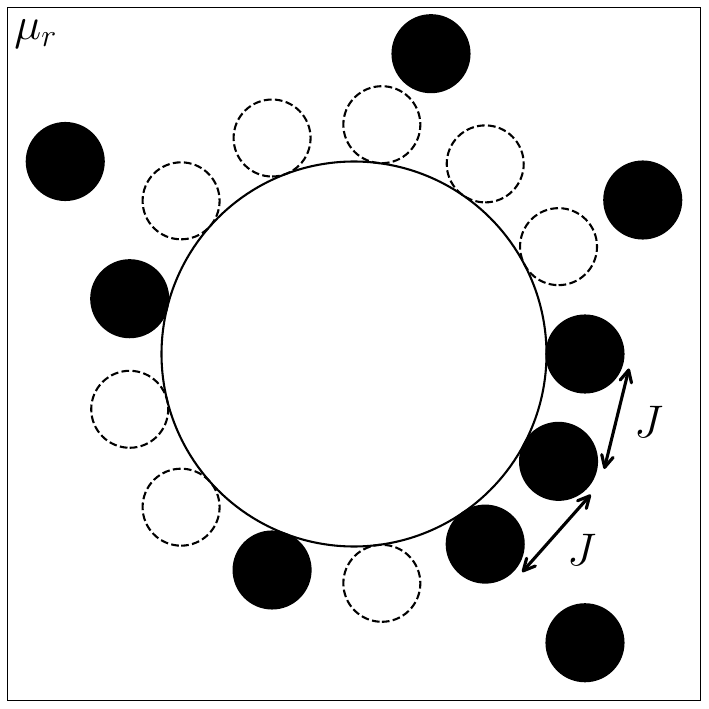}};
\end{tikzpicture}
\caption{Contour plot and heat map of the mean relative occupancy of the lattice at equilibrium, $\langle \varphi \rangle$, given by Eq.~\eqref{eq: mean relative occupancy}, as a function of the dimensionless interaction potential $J$ and chemical potential $\mu$ ranging from $-10$ to $10$, for a lattice size $L=13$ (adapted from \cite{franco2025signature}).
The insert is a schematic representation of the adsorption process, where particles bind and unbind to fixed sites. The interaction $J$ acts between bound nearest-neighbor particles, and $\mu_r$ represents the dimensionless chemical potential of the reservoir.}
\label{fig:map phi}
\end{figure}
\newline
Fig.~\ref{fig:map phi} shows the \textit{parameter diagram} of the mean relative occupancy of the lattice, $\langle \varphi \rangle$, as a function of the interaction potential $J$ and the chemical potential $\mu$.
The regions corresponding to empty and fully occupied lattices are clearly distinguishable and constitute the predominant features of the \textit{parameter diagram}. Under conditions of a negative interaction potential (repulsive nearest-neighbor interactions, hereafter termed anticooperative behavior), a significant variation of chemical potential is necessary to access different occupancy states.
Conversely, when the interaction potential is positive (attractive nearest-neighbor interactions or cooperative behavior), a small modification of the chemical potential can result in a large change across occupancy states.
This effect is particularly pronounced for strong cooperativity at half-filling (HF), where minor variations in chemical potential can induce dramatic transitions in system occupancy from empty to fully occupied states, and vice versa.
Consequently, within this regime of strong cooperativity, the system exhibits switch-like behavior characterized by abrupt occupancy transitions.

\subsection{\label{subsec: domain walls}Domain walls and number of clusters}
A cluster is defined as a domain of consecutive occupied sites. A domain wall represents a discontinuity in the occupancy of the lattice. 
Note that neither a fully occupied lattice nor an empty one presents a discontinuity.
Thus,  with the notable exception of a cluster of $L$ sites, a cluster creates two domain walls.
The number of domain walls in a given configuration can be calculated using $W(\bm{\varphi}) = 2 \sum^{L}_{i=1} \varphi_i \left(1- \varphi_{i+1}\right)$.
The mean number of domain walls at equilibrium, $\langle W \rangle$, is therefore given by $\langle W \rangle = 2 \Xi^{-1} (\partial_\mu - \partial_J) \Xi$, that is,
\begin{equation}
  \langle W \rangle = \frac{L}{e^{J}-1} \left(\tanh\left(\frac{L}{2\xi}\right) \frac{\cosh X}{\sqrt{\sinh^2X + e^{-J}}} - 1 \right).
\label{eq: domain walls}
\end{equation} 
\begin{figure}[b]
\includegraphics[width=0.5\textwidth]{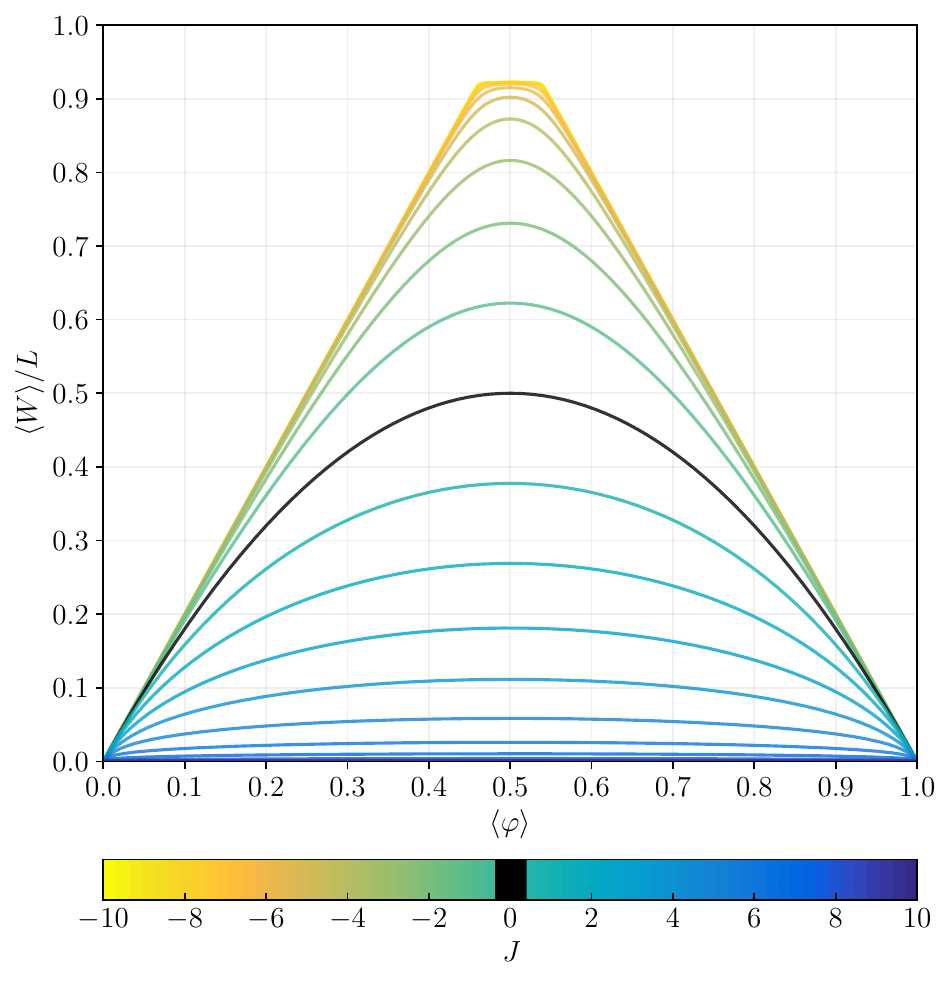}
\caption{Parametric plot of the normalized mean number of domain walls at equilibrium, $\langle W \rangle /L$, 
as a function of the mean relative occupancy $\langle \varphi \rangle$, for a lattice of size $L=13$ according to the analytical expression in Eq.~\eqref{eq: domain walls}.
For each curve,  the dimensionless interaction potential $J$ ranging from $-10$ (upmost curve) to $10$ (lowest curve) in steps of 1, is held fixed and $\mu$ is varied so that $\langle \varphi \rangle$ covers the interval $[0,1]$.
The black curve denotes the Hill-Langmuir case, $J=0$, computed from Eq.~\eqref{eq: domain walls hl}.}
\label{fig: parametric number of domain walls}
\end{figure}
The number of domain walls is symmetrical with respect to the HF line, $\langle \varphi \rangle=0.5$, as shown in Fig.~\ref{fig: parametric number of domain walls}.
In this parametric representation, each curve is obtained by fixing $J$ and varying the chemical potential $\mu$, which allows the equilibrium mean occupancy $\langle\varphi\rangle$ to sweep the full range from 0 to 1.
This plot highlights two limiting regimes of the system.
The first corresponds to a regime where the number of domain walls approaches the system size, near HF for large anticooperative interactions. 
In Fig.~\ref{fig: parametric number of domain walls}, the upper curve for large negative values of $J$ saturates near HF at a value below 1.
This is a finite-size effect: because the plot is obtained for a lattice of size $L=13$, the odd number of sites prevents the system from realizing perfect alternating particle-hole order at HF.
The second limiting regime corresponds to $\langle W \rangle=0$, which arises for strong cooperative interaction potentials independently of the occupancy, where the system behaves like a bi-stable switch between two states, either an empty or a full lattice, both of which are characterized by the absence of domain walls.

The number of clusters in a given configuration $\bm{\varphi}$ is calculated using the operator
\begin{equation}
    K(\bm{\varphi}) = \sum^{L}_{i=1} \varphi_i \left(1- \varphi_{i+1}\right) + \prod^{L}_{i=1} \varphi_i = \frac{W(\bm{\varphi})}{2} + \prod^{L}_{i=1} \varphi_i ,
\label{eq: nombre de cluster configuration locale}
\end{equation}
where the sum counts the number of final occupation sites within a given cluster of the configuration $\bm{\varphi}$ and the product accounts for the case of a full lattice, in which a single cluster of size $L$ is formed.
Hence, to find the mean number of clusters, $\langle K\rangle$, we use the mean number of domain walls in Eq.~\eqref{eq: domain walls} divided by two, as each cluster creates two domain walls, and add a correction term accounting for the case of a full system, in which there is a cluster but no domain walls.
This correction term equals the probability that the lattice is fully occupied, namely $P_\varphi(1) = e^{L(J + \mu)}/\Xi$.
Thus, the mean number of clusters can be written as:
\begin{equation}
  \langle K\rangle = \frac{\langle W\rangle}{2} + \frac{e^{2LX}}{\Xi} \, .
\label{eq: number of cluster}
\end{equation}
\begin{figure}[h]
\centering
\includegraphics[width=0.5\textwidth]{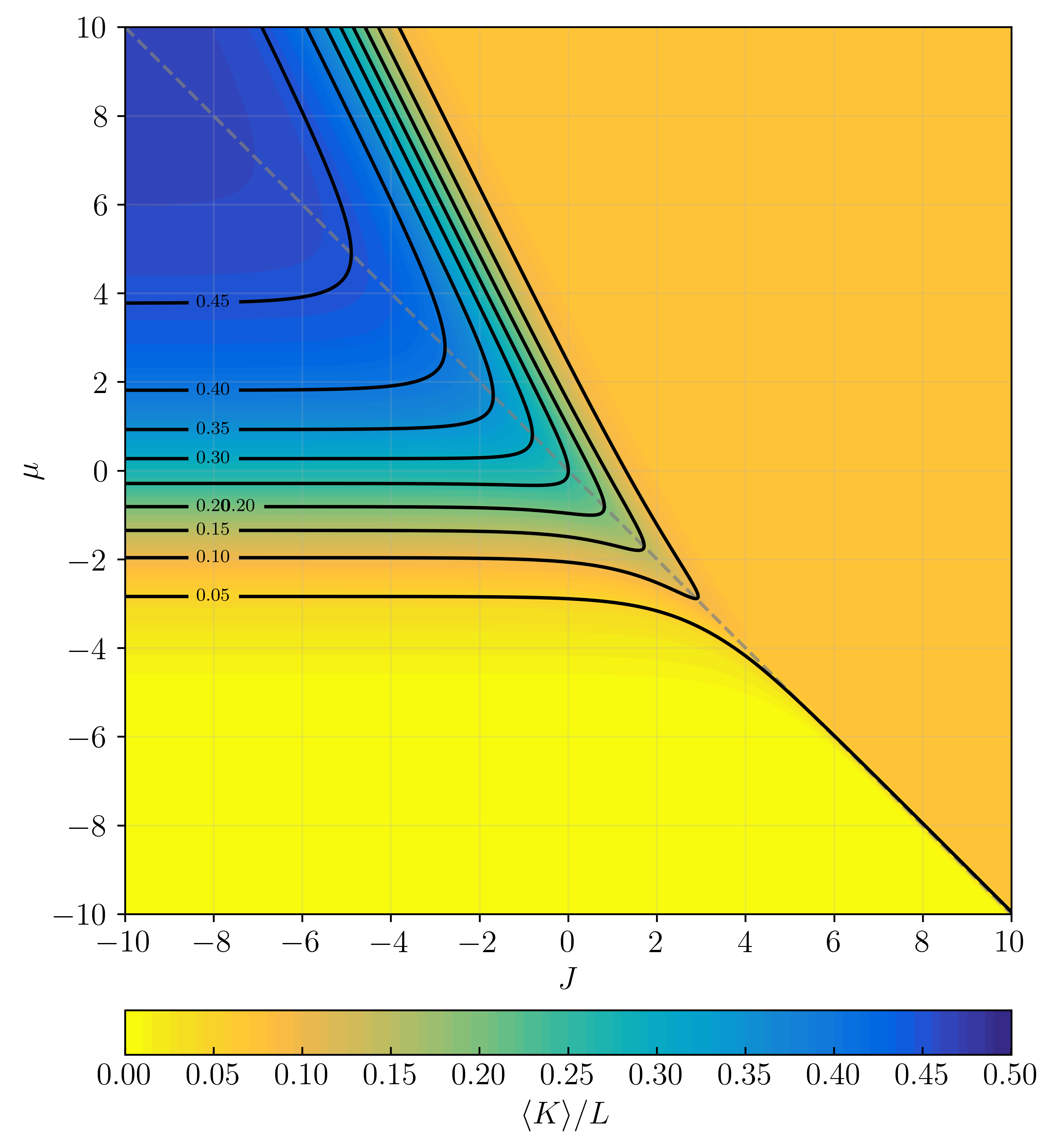}
\caption{Contour plot and heat map of the normalized mean number of clusters at equilibrium, $\langle K \rangle /L$, as a function of the dimensionless interaction potential $J$ and chemical potential $\mu$ ranging from $-10$ to $10$, according to Eq.~\eqref{eq: number of cluster} for a lattice of size $L=13$.
The dashed gray line represents the Half-Filling of the lattice, where $J+\mu=0$.}
\label{fig:map K}
\end{figure}
In the parameter diagram of the mean relative occupancy $\langle \varphi \rangle$, Fig.~\ref{fig:map phi}, the region spanning from the upper right quadrant toward the HF line is characterized by densely populated lattice configurations.
Within this region, domain walls are absent (Fig.~\ref{fig:map dw}) but a single cluster is present (Fig.~\ref{fig:map K}).
The absence of domain walls in this regime provides clear evidence for the formation of a system-sized cluster.

This asymmetry is also apparent when comparing parametric plots of $\langle W \rangle$ in Fig.~\ref{fig: parametric number of domain walls} and $\langle K \rangle$ in Fig.~\ref{fig: parametric number of cluster}.
For $J \gg 1$, the mean number of clusters varies linearly between $0$ and $1$ with the occupancy, whereas the mean number of domain walls is constant and equal to $0$ independently of the occupancy.
In this regime, the system first nucleates by attaching a particle and increases its occupancy by adding particles to the edges of the already existing cluster.
Hence, for a non-empty lattice, the dominant configurations contain a single cluster.
This regime, similar to the spin-block in the Ising model, will now be referred to as the \textit{nucleation and coalescence} regime.
The number of clusters is maximized for $J \ll -1$ at HF, where the number of bound particles is similar to the number of holes, which leads to a perfect mixing of particles and holes due to the repulsive interactions of the anticooperative system, effectively creating $L/2$ particle-hole dimers if the system is even and $(L-1)/2$ if the system is odd.
The effects of the parity of the system in limiting cases are explored in more detail in Sec.~\ref{sec: half filling}.
When looking at higher occupancy values, the number of clusters decreases as the holes are filled, fusing existing clusters until a single cluster of the system size remains, which is apparent in Fig.~\ref{fig: parametric number of cluster}.
This behavior will be referred to as the \textit{particle-hole dimer} regime.
We can also remark that the corresponding curves remain symmetric with respect to HF in both of these regimes, as well as other intermediate regimes (such as HL).
\subsection{Cluster size distributions}
We can enumerate the number $n_k$ of clusters of size $k$ in a given configuration $\bm{\varphi}$ as follows:
\begin{widetext}
\begin{equation}
    n_k(\bm{\varphi}) = \left\{
    \begin{array}{ll}
        \sum\limits^{L}_{i=1} (1-\varphi_i) \left(\prod\limits^k_{j=1} \varphi_{i+j} \right) (1 - \varphi_{i+k+1}) & \textrm{for}\;k\in\{1,\ldots, L-1\}\\
        \prod\limits^L_{i=1} \varphi_{i} & \textrm{for}\; k=L
    \end{array}
\right.
\label{eq: nk local}
\end{equation}
\end{widetext}
Here, we treat the case $k=L$ separately, since the corresponding cluster contains no domain walls owing to the periodic boundary conditions of the lattice.
The expected number $\langle n_k \rangle$ of clusters of size $k$ in the grand canonical ensemble described by $\Xi$, for fixed parameters $J$ and $\mu$, is given by:
\begin{equation}
    \langle n_k\rangle  = \frac{1}{\Xi}\sum_{\bm{\varphi}} n_k(\bm{\varphi}) e^{-\beta \mathcal{H}(\bm{\varphi})}.
    \label{eq: mean number of cluster of size k}
\end{equation}
\begin{figure}
\includegraphics[width=0.5\textwidth]{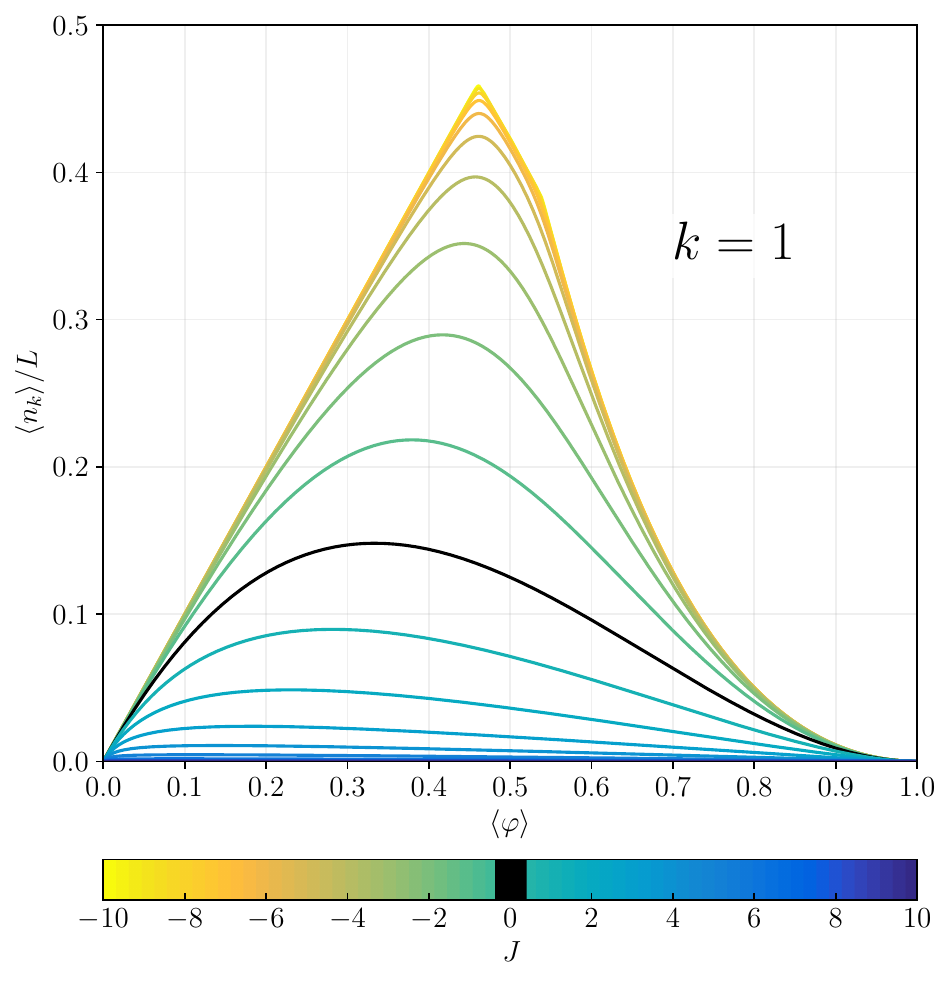}
\caption{Parametric plot of the normalized mean number of clusters of size $k$ at equilibrium, $\langle n_k \rangle /L$, as a function of the mean relative occupancy $\langle \varphi \rangle$, for a lattice of size $L=13$ and $k=1$ according to the analytical expression in Eq.~\eqref{eq: statistical weights correlations}.
For each curve,  the dimensionless interaction potential $J$ ranging from $-10$ (upmost curve) to $10$ (lowest curve) in steps of 1, is held fixed and $\mu$ is varied so that $\langle \varphi \rangle$ covers the interval $[0,1]$.
The black curve denotes the Hill-Langmuir case, $J=0$, computed from Eq.~\eqref{eq: mean number of clusters of size k hl}.}
\label{fig:parametric mean number of cluster of size k = 1}
\end{figure}
The evolution of the normalized mean number of clusters of size $k=1$ as a function of the mean relative occupancy is presented in Fig.~\ref{fig:parametric mean number of cluster of size k = 1}.
As expected, there are no clusters of size one for strongly cooperative cases.
The maximum number of clusters of size 1 is reached for strongly anticooperative cases, but parity influences not only the magnitude but also the position of the maxima.
Fig.~\ref{fig:parametric mean number of cluster of size k = 1} presents results for an odd lattice, as such, the height and coordinate of the maxima are both given by $(L-1)/2L$; in the case of an even system, both are equal to $1/2$ as the system is perfectly mixed.
At high occupancy, mono-particle clusters become less probable as the configuration-averaged cluster size increases sharply (see, for instance, Fig.~\ref{fig:parametric mean cluster size}).
At full occupancy, no mono-particle clusters remain as the system is described by a single lattice-sized cluster.

Let us now define the probability $P(k)$ of occurrence of a cluster with size $k \in \{1,\dots,L\}$. As $n_k(\bm{\varphi})$ precisely counts the number of clusters with size $k$ within the configuration $\bm{\varphi}$, $P(k)$ is obtained by first summing the statistical weights $n_k(\bm{\varphi}) e^{-\beta \mathcal{H}(\bm{\varphi})}$ over the entire set of configurations $\bm{\varphi}$ and then by normalizing the result over the entire range of sizes $k \in \{1,\dots,L\}$. Therefore, 
\begin{equation}
    P(k) = \frac{\langle n_k \rangle}{\sum\limits_{i=1}^L \langle n_i \rangle} = \frac{\langle n_k \rangle}{ \langle K\rangle} \, ,
    \label{eq: probability occurrence cluster}
\end{equation}
where the sum across all mean cluster sizes at equilibrium is simply the expected total number of clusters introduced above: $\sum_{k=1}^L \langle n_k\rangle =\langle K \rangle$.
In what follows, we refer to this probability distribution over cluster sizes as the \textit{cluster-size distribution} (CSD).
\begin{figure*}[t]
\centering

\centering
\includegraphics[width=1\textwidth]{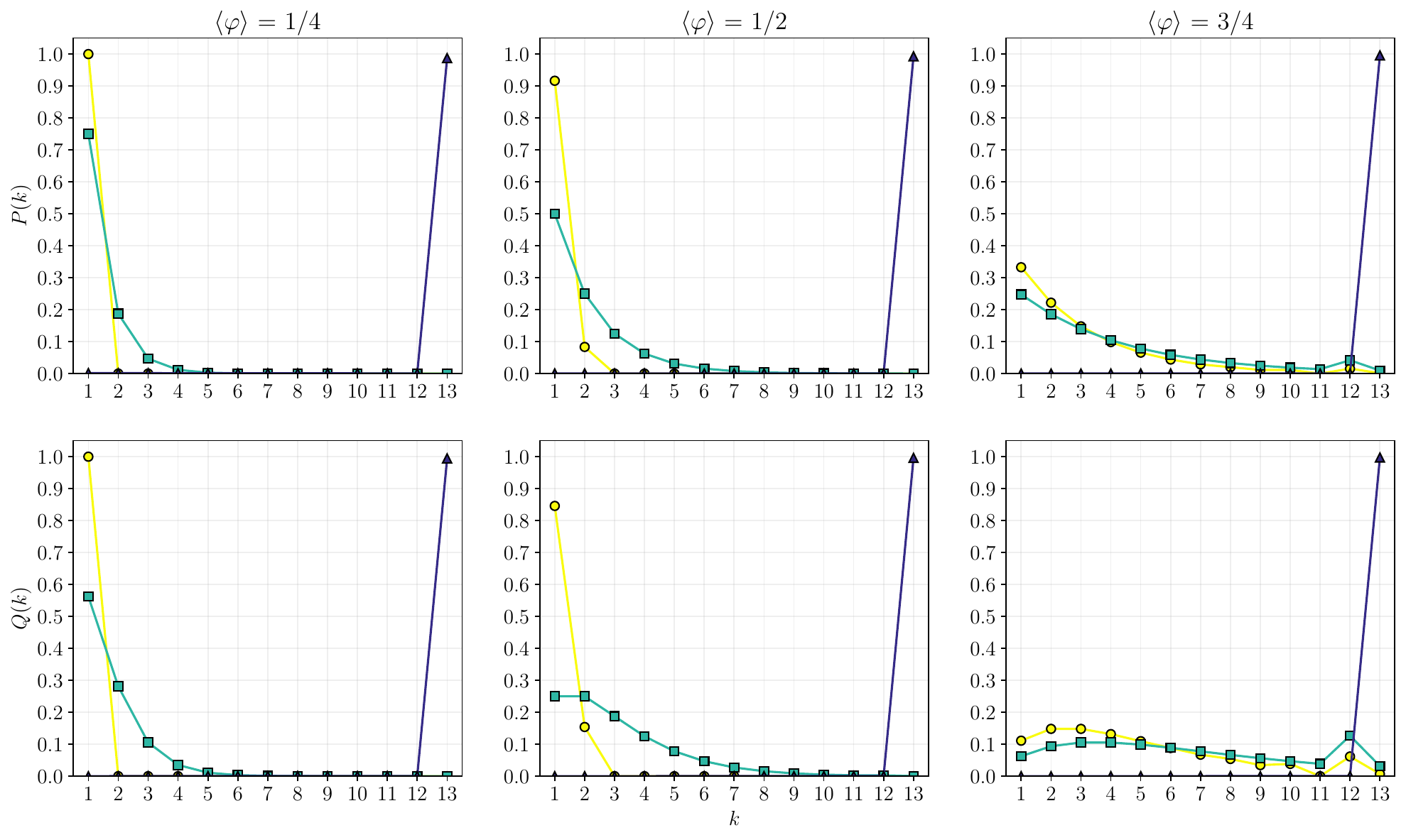}

\caption{(upper row) Cluster-size distribution, $P(k)$, calculated with Eq.~\ref{eq: probability occurrence cluster} and (lower row) site-weighted cluster-site distribution, $Q(k)$, calculated with Eq.~\eqref{eq:qk}, for different values of the dimensionless interaction potential, $J=-10$ (yellow circles), $J=10$ (purple triangles) and $J=0$ (teal squares), as well as for different values of the mean relative occupancy, $\langle \varphi \rangle = 1/4$, $1/2$ and, $3/4$.
These distributions are obtained analytically using the equilibrium expression for the number of clusters of size $k$, $\langle n_k \rangle$, given in Eq.~\eqref{eq: statistical weights correlations}.
The Hill-Langmuir case is computed using Eq.~\eqref{eq: probability occurrence cluster hl} and Eq.~\eqref{eq: qk hl} for (upper row) and (lower row), respectively.}
\label{fig:PDF_panels}
\end{figure*}
Fig.~\ref{fig:PDF_panels}(upper row) shows $P(k)$ for different values of relative occupancy and different values of the interaction potential.
In strongly cooperative systems ($J\gg 1$), the distribution approaches $P(k=L) \sim 1$ and $P(k\neq L)\sim 0$, largely independent of the mean relative occupancy chosen. 
This behavior underlies the characteristic switch-like response of such systems, where the statistical ensemble is dominated either by fully occupied or by completely empty lattice configurations.
By contrast, in strongly anticooperative systems ($J\ll -1$), single-particle clusters dominate when $\langle \varphi\rangle <0.5$, and their prevalence decreases as occupancy increases, as expected. 

The Hill-Langmuir case is discussed in detail in Sec.~\ref{sec: hill langmuir}, where explicit analytical asymptotic expressions for $P_{\mathrm{HL}}(k)$ are derived in the low- and high-occupancy regimes.
In the particular case of the HL regime at Half-Filling, the distribution probability of occurrence of a cluster of size $k$ is given by a discrete exponential decrease of the form $1/2^k$, as all cluster configurations have the same statistical weight, which is investigated in detail in Appendix~\ref{app:hfinhl} and Eq.~\eqref{eq: probability of occurrence cluster k hl hf}.
When looking outside of the HF case in the Hill-Langmuir regime, the distribution can still be approximated as a discrete exponential decrease given by $3 \langle \varphi \rangle^k$ and $(1- \langle \varphi \rangle) \langle \varphi \rangle^{k-1}$ for respectively, $\langle \varphi \rangle = 1/4$ and $\langle \varphi \rangle = 3/4$.
It is important to note that the distribution for a relative occupancy of $\langle \varphi \rangle = 3/4$, the occurrence of clusters of size $L-1$ deviates from the exponential decrease because of finite size effects.

We can also define the probability $Q(k)$ that a given site belongs to a cluster of size $k$. All sites being statistically equivalent, we show in appendix \ref{app:calculationqk} that $Q(k)$ takes the form
\begin{equation}
    Q(k) = \dfrac{ k \langle n_k\rangle}{\sum\limits_{i=1}^L i \langle n_i \rangle} = \dfrac{ k \langle n_k\rangle}{\langle N \rangle}\,.
\label{eq:qk}
\end{equation}
The weighted sum over all mean cluster sizes at equilibrium is the mean number of bound particles $\sum_{i=1}^L i \langle n_i \rangle = \langle N \rangle = L \langle \varphi \rangle$.
We will refer to this probability as the \textit{site-weighted cluster-size distribution} (SCSD).
Fig.~\ref{fig:PDF_panels} (lower row) presents the distribution $Q(k)$ for different values of the mean relative occupancy and interaction potential.
In strongly cooperative systems $(J \gg 1)$, the distribution is similar to the one observed for the CSD, where $Q(k=L) \sim 1$, and the probabilities for all other $k$ are nearly null, independently of the lattice occupancy.
In strongly anticooperative systems $(J \ll -1)$, the SCSD is also similar to the CSD, as the single-particle clusters are prevalent for occupancy $\langle \varphi \rangle < 0.5$, before becoming less important than larger-sized clusters.

While the CSD at high occupancy indicates that mono-particle clusters are the most probable configuration, the SCSD reveals that an individual site is more likely to be associated with larger clusters encompassing two, three, or four sites, rather than with a mono-particle cluster.
This difference underlines an important distinction in the interpretation of the clusters at equilibrium, as the global prevalence of mono-particle clusters contrasts with the site-weighted probability, which favors belonging to larger aggregates and highlights the coalescence of mono-particle clusters.

The Hill-Langmuir regime $(J=0)$, is described in detail in Sec.~\ref{sec: hill langmuir}.
Both distributions shown in Fig.~\ref{fig:parametric mean cluster size}, for $\langle \varphi \rangle = 1/4$ and $\langle \varphi \rangle = 3/4$, can be approximated by power laws of the form $9 k \langle \varphi \rangle^{k+1}$ and $k \langle \varphi \rangle^k/12$, respectively.
One should note that, in this regime, the distribution also exhibits a finite-size effect, as $Q(k=L-1)$ deviates from the expected power-law behavior at high occupancy.
At Half-filling, the distribution follows a power law (see Eq.~\eqref{eq:q k hl hf}) of the form $k \langle \varphi \rangle^{k+1}$, where $Q(k=1) = Q(k=2)=0.25$, meaning that a random site chosen uniformly in the lattice has the same probability of being in a mono-particle cluster or a dimer cluster.
Interestingly, this behavior is not parity-dependent.
\newline
The expectation value of the cluster size with respect to the CSD, $P(k)$,
\begin{equation}
    \kappa = \sum^{L}_{k=1} k P(k) = \frac{\sum\limits_{k=1}^L k \langle
     n_k \rangle}{\sum\limits_{k=1}^L \langle
     n_k \rangle} \, ,
    \label{eq: average cluster size}
\end{equation}
represents the average number of sites in a \textit{cluster chosen} uniformly at random from the statistical ensemble of lattice configurations defined by $\Xi$.
This can also be understood as the typical or most probable cluster size to be encountered for a given set of parameters $J$ and $\mu$.
Note that the nominator in the above expression is just the expected number of occupied sites and the denominator is the expected number of clusters. Therefore, we find:
\begin{equation}
    \kappa = \frac{L\langle \varphi \rangle}{\langle K\rangle}
    \label{eq: kappa}
\end{equation}
which captures the intuitive idea that the mean occupancy is proportional to the product of the average cluster size and the average number of clusters.

\begin{figure}[t]
\includegraphics[width=0.5\textwidth]{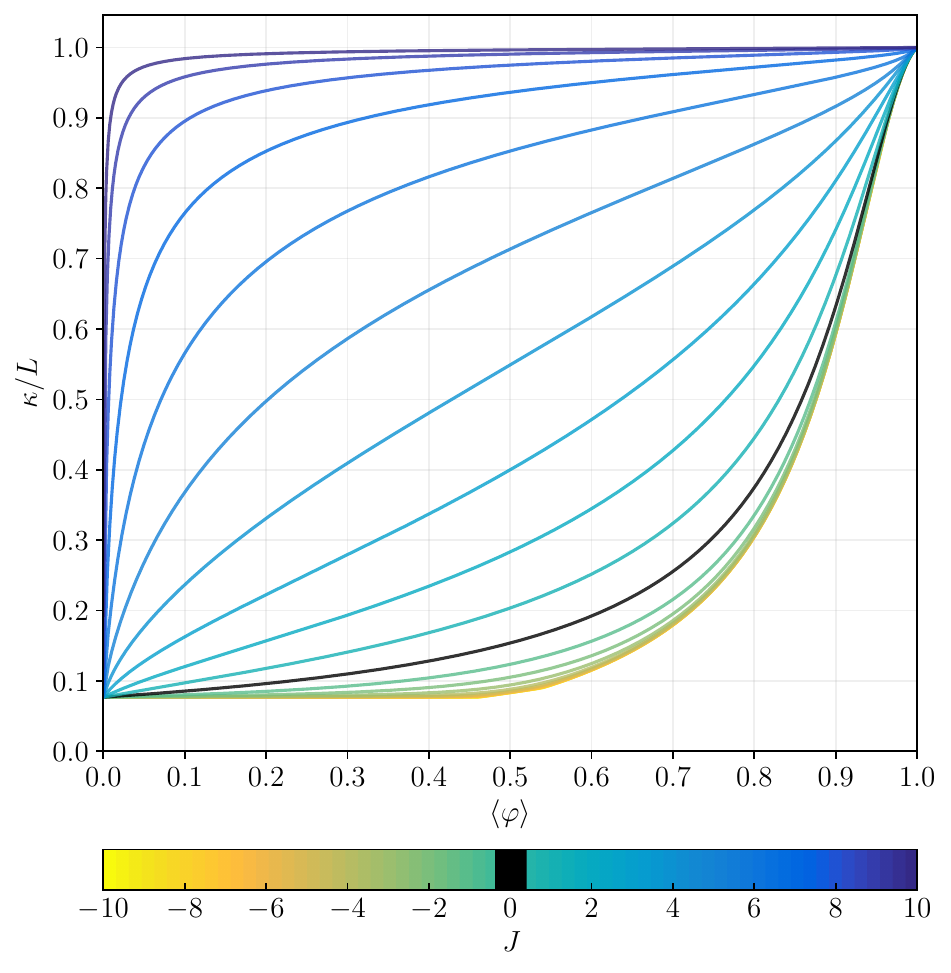}
\caption{Parametric plot of the normalized average cluster size at equilibrium, as a function of the mean relative occupancy $\langle \varphi \rangle$, for a lattice of size $L=13$ according to the analytical expression in Eq.~\eqref{eq: kappa}.
For each curve,  the dimensionless interaction potential $J$ ranging from $-10$ (lowest curve) to $10$ (upmost curve) in steps of 1, is held fixed and $\mu$ is varied so that $\langle \varphi \rangle$ covers the interval $[0,1]$.
The black curve denotes the Hill-Langmuir case, $J=0$, computed from Eq.~\eqref{eq: average cluster size hl}.}
\label{fig:parametric average cluster size}
\end{figure}

The switch-like behavior of strongly cooperative systems is apparent in Fig.~\ref{fig:parametric average cluster size}: for any nonzero occupancy, the most probable cluster size is the system size $L$. 
By contrast, the particle-hole dimer regime is also clearly identifiable in the anticooperative case ($J>0$): here, the curves resemble those found for the corresponding configuration-averaged cluster size, as shown in Fig.~\ref{fig:parametric mean cluster size}.
At low particle densities, the plateau where the number of particles equals the number of clusters indicates that the particles are predominantly isolated. This regime is followed by a sharp rise as larger clusters begin to form.
\subsection{$k$-site cluster correlation functions}
The quantities introduced in the sections above, such as the mean number of clusters of size $\langle K \rangle$, the CSD, and the SCSD, were calculated using exact enumeration. 
To calculate them analytically, it is convenient to introduce the cluster correlation function of size $k$ defined as:
\begin{equation}
  c_k = \frac{1}{\Xi} \sum_{\bm{\varphi}} \left(\prod_{i=1}^k \varphi_i \right) e^{-\beta \mathcal{H}(\bm{\varphi})} \, ,
  \label{eq: ck}
\end{equation}
As the system is invariant under lattice-site translations, a correlation function $c_k$ depends only on the length $k$ of the cluster, and not on its position.
Using Eq.~\eqref{eq: nk local} and Eq.~\eqref{eq: mean number of cluster of size k}, we can express the mean number of clusters of size $k$ in terms of the cluster correlation functions as:
\begin{equation}
  \langle n_k \rangle  = \left\{
    \begin{array}{ll}
        L(c_k - 2 c_{k+1} + c_{k+2}) &, \forall k \in \{1,...,L-2\} \\
        L(c_{L-1} - c_L) &, k=L-1 \\
        c_L &, k=L
    \end{array}
\right.
\label{eq: statistical weights correlations}
\end{equation}
We can also express the correlation function using the transfer matrix $T$ as: 
\begin{equation}
  c_k = \frac{e^{k\mu + J (k-1)}}{\Xi} \langle u|T^{L-k-1}|u\rangle \, ,
  \label{eq: ck transfer matrix}
\end{equation}
with
\begin{equation}
  |u\rangle = 
  \begin{pmatrix}
  1 \\
  e^{J + \frac{\mu}{2}}
  \end{pmatrix}.
\end{equation}
Finally, we obtain an explicit expression for the cluster correlation function in terms of the eigenvalues of $T$:
\begin{equation}
  c_k = \frac{e^{k\mu + J(k-1)}}{\Xi} \left( u_+ \lambda_+^{L-k-1} + u_- \lambda_-^{L-k-1} \right) \, ,
  \label{eq: ck eigenvalues}
\end{equation}
where 
\begin{equation}
  u_\pm = \frac{e^\mu (1 + e^J (\lambda_\pm -1))^2}{e^\mu + (\lambda_\pm - 1)^2}\, .
  \label{eq: u pm}
\end{equation}
We can then use the correlation functions to redefine simple quantities, such as the mean relative occupancy, which is $\langle \varphi \rangle = \langle \varphi_1 \rangle = c_1$. Using the rotational invariance of these functions, we rewrite the mean number of clusters as:
\begin{equation}
  \langle K \rangle = L \left( \langle \varphi_1 \rangle - \langle \varphi_1 \varphi_2 \rangle \right) + \langle \prod_{i=1}^{L} \varphi_i \rangle\,,
  \label{eq: number of clusters phi}
\end{equation}
which immediately gives:
\begin{equation}
  \langle K \rangle = L(c_1 - c_2) + c_L\,.
  \label{eq: number of clusters correlation}
\end{equation}
Using this expression, we can derive a formula for $\kappa$ with the same correlation functions
\begin{equation}
  \kappa = \frac{L c_1}{L \left(c_1 - c_2 \right) + c_L}\,,
  \label{eq: average cluster size correlation}
\end{equation}
where $L c_1$ is the mean number of bound particles, $\langle N \rangle$.
The $k$-site cluster correlation functions allow us to give an exact definition of the mean number of cluster of size $k$, which, consequently, gives us access to exact expressions for both distributions $P(k)$ and $Q(k)$, and the expectation value of the cluster size $\kappa$, without having to employ the enumeration method over all possible configurations.
\subsection{Configuration-averaged cluster size}
The average cluster size $C$ in a given non-empty lattice configuration $\bm{\varphi}$ is defined as the ratio of the number of bound particles and the number of clusters: $ N(\bm{\varphi})/K(\bm{\varphi})$.
We denote the expectation value of this observable with respect to the grand canonical probability distribution over microstates,
\begin{equation}
    \langle C \rangle = \frac{1}{\Xi} \sum_{\bm{\varphi} \neq \bm{\varphi_0}} \frac{N(\bm{\varphi})}{K(\bm{\varphi})}e^{-\beta \mathcal{H}(\bm{\varphi})}\,,
    \label{eq: mean cluster size}
\end{equation}
as the \textit{configuration-averaged cluster size}, where $\bm{\varphi_0}$ represents the empty configuration.
In order to calculate this quantity analytically, we performed the variable change $\mu = 2 X - J$ in the partition function, which becomes:
\begin{align}
    \Xi(J,\mu &=2X-J) = \nonumber\\ 
    &=\sum_{\bm{\varphi}}  \,\exp\left[ - J \sum_{i=1}^L \varphi_i(1 - \varphi_{i+1}) + 2 X \sum_{i=1}^L \varphi_i \right] \nonumber\\
    & = 1 + e^{2LX} + 
    \hat{\Xi}(J,X) \,.
\label{eq: partition function var change}
\end{align}
The first and second terms in the last equation correspond to the contributions of the empty lattice, $\beta\mathcal{H}(\boldsymbol{\varphi}_0)=0$, and the fully occupied lattice, $\beta\mathcal{H}(\boldsymbol{\varphi}_1)=2L X$, respectively, while the third term represents the reduced partition function:
\begin{equation}
    \hat{\Xi}(J,X) = \sum_{\bm{\varphi} \neq (\bm{\varphi}_0, \bm{\varphi}_1)} e^{-JK(\bm{\varphi}) + 2 X N (\bm{\varphi})},
\label{eq: reduced partition function}
\end{equation}
where we have identified the operators for the number of clusters $K(\bm{\varphi})$ and number of particles $N(\bm{\varphi})$ in the exponential term of the Boltzmann weights in Eq.~\eqref{eq: partition function var change}.
This allows us to rewrite $\langle C \rangle$ as:
\begin{equation}
    \langle C \rangle = \frac{1}{\Xi}\left(Le^{2LX} + \frac{1}{2}\frac{\partial}{\partial X} \int_J^\infty \hat{\Xi}(J',X) dJ' \right)\,.
\label{eq: integral mean cluster size}
\end{equation}
Upon expanding the eigenvalues in Eq.~\eqref{eq: eigenvalues} and evaluating the integral over $J'$ in Eq.~\eqref{eq: integral mean cluster size}, we eventually find an explicit expression for the configuration-averaged cluster size that reads:
\begin{widetext}
\begin{equation}
\langle C \rangle = \frac{1}{\Xi} \left[Le^{2LX} + \frac{\partial }{\partial X} \left( e^{LX} \sum_{k=1}^{\left\lfloor \frac{L}{2}\right\rfloor} \sum_{n=1}^k \binom{L}{2k} \binom{k}{n} \frac{1}{n} \cosh(X)^{L-2k} \sinh(X)^{2(k-n)} e^{-nJ} \right) \right]\,.
\label{eq: mean cluster size exact}
\end{equation}
\end{widetext}

\begin{figure}
\includegraphics[width=0.5\textwidth]{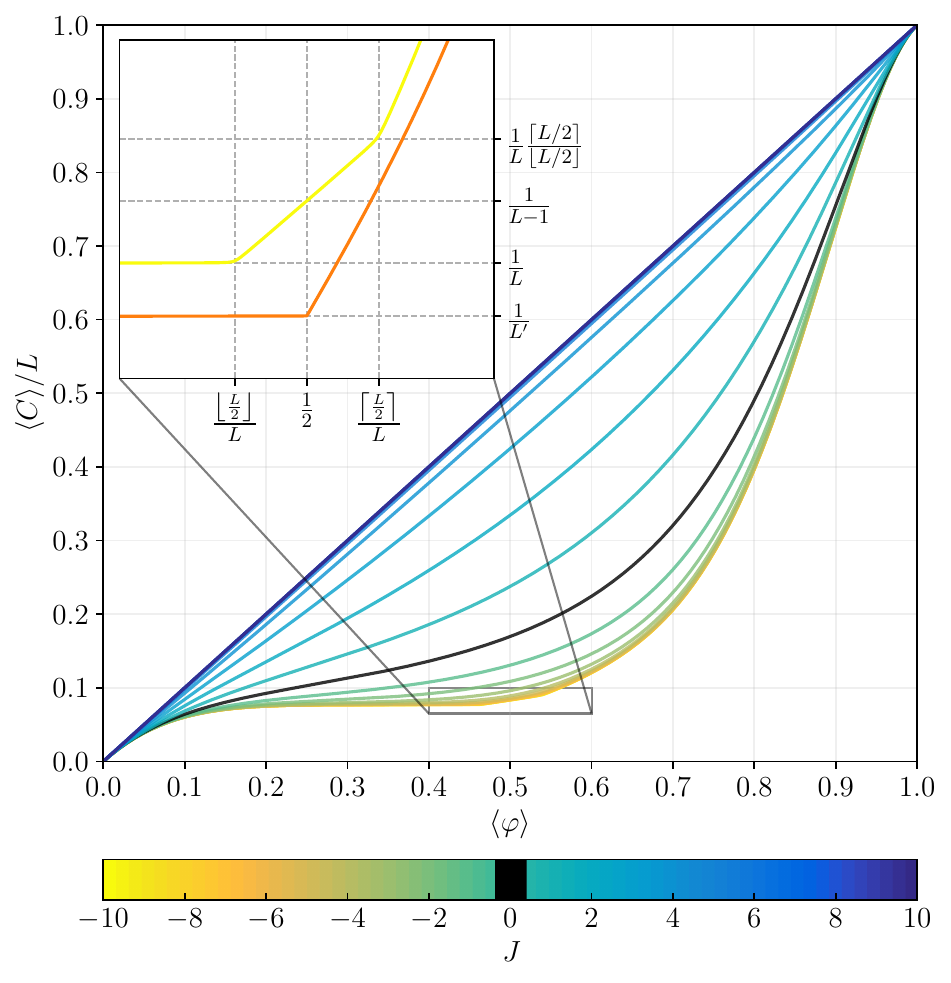}
\caption{Parametric plot of the normalized configuration-averaged cluster size at equilibrium, $\langle C \rangle /L$, as a function of the mean relative occupancy $\langle \varphi \rangle$, for a lattice of size $L=13$ according to the analytical expression in Eq.~\eqref{eq: mean cluster size}.
For each curve,  the dimensionless interaction potential $J$ ranging from $-10$ (upmost curve) to $10$ (lowest curve) in steps of 1, is held fixed and $\mu$ is varied so that $\langle \varphi \rangle$ covers the interval $[0,1]$.
The black curve denotes the Hill-Langmuir case, $J=0$, computed from Eq.~\eqref{eq: mean cluster size hl}.
The inset represents the evolution $\langle C \rangle /L$ in the anticooperative limit ($J=-10$) for $L=13$ in yellow and for $L^\prime = 14$ in orange, as a function of $\langle \varphi \rangle$ for $\langle \varphi \rangle \in [0.4,0.6]$.}
\label{fig:parametric mean cluster size}
\end{figure}

As shown in Fig.~\ref{fig:parametric mean cluster size}, in the strongly cooperative cases, the normalized configuration-averaged cluster size, $\langle C \rangle /L$, increases linearly with the average occupancy meaning that once the first particle is attached, the next will always bound to the edges of the previously formed cluster, hence the number of particles increases while the number of cluster remains constant.
This can be proven by evaluating the partition function, the mean relative occupancy and the configuration-averaged cluster size for $J\to \infty$ with a fixed $J$:
\begin{equation}
    \Xi = \lambda_+^L + \lambda_-^L \underset{J\to \infty}{\to} 1 + e^{2LX}\,,
    \label{eq: partition function strong coop}
\end{equation}
where these terms represent the Boltzmann weight of the two states participating in the switch-like behavior, \textit{i.e.} an empty and full configuration.
The mean relative occupancy is found via the formula $\langle \varphi \rangle = (2L\Xi)^{-1} \partial_X \Xi$, which gives
\begin{equation}
    \langle \varphi \rangle \underset{J\to \infty}{\to} \frac{e^{2LX}}{1 + e^{2LX}}\,,
    \label{eq: mean relative occupancy strong coop}
\end{equation}
and thanks to Eq.~\eqref{eq: mean cluster size} we find:
\begin{equation}
    \langle C \rangle \underset{J\to \infty}{\to} \frac{L e^{2LX}}{1 + e^{2LX}}\,.
\end{equation}
Therefore, in this highly cooperative limit, at $X$ fixed, we have:
\begin{equation}
    \frac{\langle C \rangle}{L} \underset{J\to \infty}{=} \langle \varphi \rangle\,.
\label{eq: mean cluster size strong coop}
\end{equation}

In the strongly anticooperative limit, the relationship between $\langle C \rangle /L$ and $\langle \varphi \rangle$ is less straightforward than the one obtained in the last paragraph, as this limit presents three different regimes.
In the low density regime, for occupancy in the range  $\langle \varphi \rangle \in \left[0, \lfloor L/2 \rfloor /L \right]$, particles strongly repel each other and are naturally separated by an empty site. 
In other words, each particle added to the system also adds a cluster meaning that the configuration-averaged cluster size remains constant as $N(\bm{\varphi}) = K(\bm{\varphi})$, resulting in the plateau observed in Fig.~\ref{fig:parametric mean cluster size}.
This can be characterized by studying the limit $J\to -\infty$ with $\mu = -J+y$ where $y$ is an independent variable which can be expressed in terms of $X$ by $y = 2X$.
By introducing the fugacity,
\begin{equation}
    z = e^\mu \in [0,\infty[\,,
    \label{eq: fugacity}
\end{equation}
the partition function, in this limit,  becomes
\begin{equation}
    \Xi = \sum^{\left\lfloor \frac{L}{2} \right\rfloor}_{k=0} \frac{L}{L-k} \binom{L-k}{k} z^k\, ,
\end{equation}
and the normalized configuration-averaged cluster size reads:
\begin{equation}
    \frac{\langle C \rangle}{L} = \frac{1}{\Xi} \sum^{\left\lfloor \frac{L}{2} \right\rfloor}_{k=1} \frac{1}{L-k} \binom{L-k}{k} z^k \, .
\end{equation}
Hence, we can calculate the limits of this observable at the boundaries of occupancy in this regime leading to:
\begin{equation}
    \lim_{z\to0} \frac{\langle C \rangle}{L} = 0 \quad \text{and} \quad 
    \lim_{z\to\infty} \frac{\langle C \rangle}{L} = \frac{1}{L}\,.
\label{eq: lim mean cluster size low density}
\end{equation}
In the limited occupancy range \mbox{$\langle \varphi \rangle \in \left[ \lfloor L/2 \rfloor /L ,\lceil L/2 \rceil /L\right]$}, the system is in a geometrically frustrated regime (cf.  inset in Fig.~\ref{fig:parametric mean cluster size}).
Using an Ising analogy, at the lower bound of this regime the system can be reduced to the following pattern $\{\downarrow \overbrace{ \uparrow \downarrow\downarrow}\uparrow\downarrow \}$ where the pair $\{\downarrow \uparrow \}$ represents the particle-hole dimer that fills the lattice as particles strongly repel each other, as seen in the low density regime.
In the corresponding configurations, $\langle C\rangle = 1$ as seen in Eq.~\eqref{eq: lim mean cluster size low density}. 
The upper bound is represented by $\{\downarrow \overbrace{ \uparrow \uparrow\downarrow}\uparrow\downarrow \}$ where the geometrical frustration caused by the triplet is apparent.
It should be noted that when the lattice size is even, this behavior reduces to a single point, namely $\langle \varphi \rangle = 1/2 $, with $\langle C \rangle = 1$ reflecting the perfect mixing of particles.
The analytical expression is obtained by taking the limit of $J \to - \infty$ with $\mu = - J + y$, where $y$ is fixed. 
The configuration-averaged cluster size depends linearly on the mean relative occupancy as 
\begin{equation}
    \frac{\langle C \rangle}{L} = \frac{2}{L-1} \langle \varphi \rangle,
    \label{eq: mean cluster size frustrated}
\end{equation}
thus for an odd system in this regime, the range is $ \frac{\langle C \rangle}{L} \in \left[ \frac{1}{L}, \frac{1}{L}\frac{L+1}{L-1} \right]$.

Finally, the high-density regime in the strongly anticooperative limit is identified for mean relative occupancy comprised between $\langle \varphi \rangle \in \left[ \lceil L/2 \rceil /L ,1\right]$.
In this regime, the normalized configuration-averaged cluster size shows sharp variations as each added particle fills an empty site between two particles, hence $N(\bm{\varphi}) > K(\bm{\varphi})$ until $C(\bm{\varphi}) = N(\bm{\varphi}) = L$ as a full system is constituted of a single $L$-sized cluster.
To study the limit of this regime we set $J \to - \infty$ with $\mu = -2J + y$ fixed and define the auxiliary variable
\begin{equation}
    x = e^y \in [0,\infty[\,.
\end{equation}
Then the mean configuration-averaged cluster size can be written as 
\begin{equation}
    \frac{\langle C \rangle }{L} = \frac{x^L + \sum\limits_{k=1}^{\left\lfloor \frac{L}{2}\right\rfloor} \frac{1}{k} \binom{L-k}{k} x^{L-k}}
    {x^L + \sum\limits_{k=1}^{\left\lfloor \frac{L}{2}\right\rfloor} \frac{L}{L-k} \binom{L-k}{k} x^{L-k}}\,,
\label{eq: lim mean cluster size high density}
\end{equation}
which gives the following range in this regime: $\frac{\langle C \rangle}{L} \in \left[ \frac{1}{L} \frac{\left\lceil \frac{L}{2}\right\rceil}{\left\lfloor \frac{L}{2}\right\rfloor} ,1\right]$.
\section{\label{sec:cluster formalism}Cluster formalism}
In this section, we provide exact results for large lattices, for which an exact enumeration of microscopic lattice configurations becomes computationally prohibitive.
We aim to compute the same quantities as those introduced in Sec.~\ref{sec: occupation formalism}, namely the partition function and the associated average cluster observables.
Here, we achieve this by applying a combinatorial approach that reduces the configuration space by characterizing each configuration solely through the number and sizes of its clusters, thereby allowing the application of exact enumeration techniques to larger systems than those available in the previous section.

To characterize the cluster assembly, the partition function needs to be rewritten. Adapting the results of \textcite{yilmaz2005exact}, the Hamiltonian describing the energy of the system as a function of a cluster configuration $\bm{q}$ for a one-dimensional lattice of finite size $L$ in the SRLG with periodic boundary conditions can be written as:
\begin{equation}
  \beta \mathcal{H}(\bm{q}) = -J \sum^{L}_{k=1} (k-1+\delta_{k,L})q_k(\bm{q}) - \mu \sum^{L}_{k=1} k q_k(\bm{q})\,.
\label{eq: hamiltonian cluster}
\end{equation}
The vector $\bm{q}$ represents a possible cluster configuration, satisfying the conditions:
\begin{equation}
  N(\bm{q}) = \sum^{L}_{k=1} k q_k(\bm{q})\,,
\label{eq: condition n}
\end{equation}
and 
\begin{equation}
  K(\bm{q}) = \sum^{L}_{k=1} q_k(\bm{q})\,.
\label{eq: condition q}
\end{equation}
Where $N(\bm{q})$ is the number of bound particles and $K(\bm{q})$ is the number of clusters in the configuration $\bm{q} = \{q_k\} = \{q_1, q_2, ..., q_L \}$.
In this notation, the variable $q_k$ represents the number of clusters of size $k$ in a configuration $\bm{q}$.
The partition function describing the exact cluster size distribution is:
\begin{equation}
  \Xi = \sum_{\bm{q} \in \mathcal{Q}} g(\bm{q}) e^{-\beta \mathcal{H}(\bm{q})}\,.
\label{eq: partition function cluster}
\end{equation}
The sum over $\bm{q} \in \mathcal{Q}$ represents the sum over all possible cluster configurations satisfying Eqs.~\eqref{eq: condition n} and \eqref{eq: condition q}.
The degeneracy of a cluster configuration is defined by the function $g(\bm{q})$ and is given in Eq.~\eqref{eq: degeneracy cluster} by adapting to the finite the lattice-gas framework the degeneracy function originally proposed by \textcite{yilmaz2005exact} for the one-dimensional Ising model:
\begin{widetext}
\begin{equation}
  g(\bm{q}) = \frac{L \left( L - \sum\limits_{k=1}^{L} (k - \delta_{k,L}) q_k(\bm{q}) - 1 \right)!}{(q_L(\bm{q})(L-1) + 1) \left( L - \sum\limits_{k=1}^{L} (k + 1 - \delta_{k,L}) q_k(\bm{q}) \right)! \prod\limits_{k=1}^{L} q_k(\bm{q})!}\,.
\label{eq: degeneracy cluster}
\end{equation}
\end{widetext}
The set $\mathcal{Q}$ contains vectors of size $L$ representing all possible cluster configurations for a one-dimensional lattice of finite dimension and is defined as: 
\begin{equation}
  \mathcal{Q}(L) = \{\bm{q} \in \mathbb{N}^{L} | N(\bm{q}) + K(\bm{q}) - q_L(\bm{q}) \leqslant L \}\,,
\label{eq: set Q}
\end{equation}
where $\mathbb{N}$ denotes the set of non-negative integers (including zero). 
In the cluster formalism, we use exact enumeration to calculate the mean values at equilibrium of the previously established observables.
The cluster formalism is particularly well suited for this method as the cardinality of the set $\mathcal{Q}(L)$ scales as $p(L)+1$, where $p(L)$ is the integer partition of $L$, which is always inferior to the $2^L$ configurations in the lattice description of the system, especially for large lattice sizes \cite{comtet2012advanced}.
\subsection{\label{subsec: statistical weights}Statistical weights}
We define the statistical weights, $w_k$, by:
\begin{equation}
  w_k = \sum_{\bm{q} \in \mathcal{Q}} q_k(\bm{q}) g(\bm{q}) e^{-\beta \mathcal{H}(\bm{q})}\,.
\label{eq: cf statistical weights}
\end{equation}
These statistical weights represent the unnormalized contribution of clusters of size $k$, obtained by summing $q_k(\bm{q})$ over all admissible configurations $\bm{q}\in\mathcal{Q}$.
Upon normalization by the partition function $\Xi$, they yield the mean number of clusters of size $k$, as shown in Eq.~\eqref{eq: cf mean number cluster size k}.
\subsection{\label{subsec: domain walls and clusters}Domain walls and clusters}
In the cluster formalism, the number of domain walls in a configuration $\bm{q}$ is given by $W(\bm{q})= 2 \sum_{k=1}^{L-1} q_k(\bm{q})$. The superior limit of the sum is $L-1$ as the last element of the vector, $q_L(\bm{q})$, represents a cluster of the system size which has no domain walls, hence should be excluded from the calculation.
The average number of domain walls can be computed using the statistical weights defined in Eq.~\eqref{eq: cf statistical weights}, as follows:
\begin{equation}
  \langle W \rangle = \frac{2}{\Xi} \sum_{k=1}^{L-1} w_k \,.
  \label{eq: cf domain walls}
\end{equation}
The formula for the mean number of clusters at equilibrium is even more straightforward, using the definition of the number of clusters in a configuration $\bm{q}$ given in Eq.~\eqref{eq: condition q}, we find:
\begin{equation}
  \langle K \rangle = \frac{1}{\Xi} \sum_{k=1}^{L} w_k \,.
  \label{eq: cf mean number of clusters}
\end{equation}

\subsection{\label{subsec: mean cluster size}Configuration-averaged cluster size}
The configuration-averaged cluster size in a cluster configuration $\bm{q}$ is defined as $C(\bm{q})= N(\bm{q})/K(\bm{q})$, using Eqs.~\eqref{eq: condition n} and \eqref{eq: condition q}. Multiplying this local configuration-averaged cluster size by the degeneracy and the Boltzmann weight of the corresponding configuration, then summing over all possible configurations and normalizing by the partition function, gives the configuration-averaged cluster size at equilibrium in the system:
\begin{equation}
    \langle C \rangle  = \frac{1}{\Xi} \sum_{\bm{q} \in \mathcal{Q}} C(\bm{q}) g(\bm{q}) e^{-\beta \mathcal{H}(\bm{q})} \,.
    \label{eq: cf mean cluster size}
\end{equation}
\subsection{\label{subsec: cf cluster size distribution}Cluster size distribution}
As the vector $\bm{q}$ contains the number of clusters of size $k$, the statistical weights are particularly well suited to describe the cluster statistics.
Hence, the mean number of clusters of size $k$ is written as:
\begin{equation}
    \langle n_k \rangle  = \frac{1}{\Xi} \sum_{\bm{q} \in \mathcal{Q}} q_k (\bm{q}) g(\bm{q}) e^{-\beta \mathcal{H}(\bm{q})} = \frac{w_k}{\Xi} \,.
    \label{eq: cf mean number cluster size k}
\end{equation}
The probability of occurrence of a cluster of size $k$, $P(k)$, is also define using the statistical weights:
\begin{equation}
    P (k) = \frac{w_k}{\sum\limits_{k=1}^L w_k } = \frac{w_k}{\Xi \langle K\rangle}\,.
    \label{eq: cf probability occurence cluster size k}
\end{equation}
The probability that a site belongs to a cluster of size $k$, $Q(k)$, can also be written using the statistical weights:
\begin{equation}
    Q(k)  = \frac{k w_k}{\Xi \langle N \rangle} \,.
    \label{eq: cf cluster partition function}
\end{equation}
These two definitions are identical to the one found in Eq.~\eqref{eq: probability occurrence cluster} and Eq.~\eqref{eq:qk}.
Using the definition of the probability of occurrence of a cluster of size $k$, we can define the most probable cluster size similarly as in Eq.~\eqref{eq: average cluster size} or using the statistical weights, which gives:
\begin{equation}
    \kappa = \sum_{k=1}^L k P(k) = \frac{1}{\Xi \langle K\rangle} \sum_{k=1}^L k w_k \,.
    \label{eq: cf average cluster size}
\end{equation}
\section{\label{sec: conclusion}Conclusion and discussion}
We obtained exact finite-size results for equilibrium cluster statistics in a one-dimensional short-range lattice gas with periodic boundary conditions, coupled to a single grand-canonical (thermochemical) reservoir at temperature $T$ and chemical potential $\mu$.
In particular, we obtained explicit expressions for the mean occupancy, the mean number of domain walls, the mean number of clusters, and for cluster-size observables including the mean number of clusters of size $k$ and the associated cluster-size distributions.
In addition, we introduced a complementary cluster-based combinatorial formulation in which configurations are classified by cluster counts and sizes, providing a reorganization of the partition function. As a practical consequence, the enumeration involves a set of cluster configurations whose cardinality scales as $p(L)+1$, where $p(L)$ is the number of integer partitions of $L$, rather than the $2^L$ site microstates, while still reproducing the cluster statistics exactly.
This study aims to better characterize systems where nearest-neighbor interactions shape spatial organization on ring-like substrates, with potential applications to the assembly of molecular subunits in biology, including torque-generating unit binding in the bacterial flagellar motor \cite{franco2025signature,perez2022relaxation}, and cooperative ligand binding to ring-shaped oligomeric proteins \cite{li2022thermodynamic}.
Finally, an important outcome of our exact finite-size results is the set of functional relations they imply between experimentally accessible observables. For finite rings, cluster summary statistics such as the mean number of clusters $\langle K \rangle$, the mean number of domain walls $\langle W \rangle$, and mean cluster-size measures (e.g., $\langle C \rangle$ or $\kappa$) trace distinct, $J$-dependent parametric curves when plotted against the mean occupancy $\langle\varphi\rangle$ (see Figs.~\ref{fig: parametric number of domain walls}, \ref{fig:parametric average cluster size}, \ref{fig:parametric mean cluster size}, and \ref{fig: parametric number of cluster}), providing stringent benchmarks for small periodic systems. 
This suggests a practical strategy: joint measurements of occupancy together with a single cluster observable can, in principle, constrain the nearest-neighbor coupling $J$ more tightly than occupancy-only summary statistics, complementing approaches that infer (anti)cooperativity from occupancy fluctuations alone \cite{franco2025signature}.
\begin{acknowledgments}
Discussions with A. Parmeggiani, L. Remini, and F. Geniet are gratefully acknowledged by TA.
TA, N-OW, and JP were financially supported by the French Agence Nationale de La Recherche (ANR) (BaElMec project Grant No. ANR-23-CE30-0010). 
N-OW and JP were supported by the LabEx NUMEV (ANR-10-LABX-0020) within the I-Site MUSE (ANR-16-IDEX-0006).
All authors were supported by the Centre Nationale de Recherche Scientifique (CNRS) and the University of Montpellier.
\end{acknowledgments}

\appendix

\section{\label{app:calculationqk}Probability $Q(k)$}

The probability $Q(k)$ is the conditional probability that a given site belongs to a cluster of size $k$ knowing that this site is part of a cluster.
In this sense, let us define the probability $P_r(k)$, that a site of the lattice, say site 1 for they are all equivalent, belongs to a cluster of size $k$
\begin{equation}
    P_r(k) = k \langle (1 - \varphi_L) (1 - \varphi_{k+1}) \prod_{j=1}^k \varphi_j \rangle, \quad k \in [1,L-1],
\end{equation}
for $k=L$ this simplifies to $P_r(k=L) = \langle \prod\limits_{j=1}^L \varphi_j \rangle $ as it corresponds to the fully occupied lattice.
We can then inject the $k$-site correlation function (Eq.~\eqref{eq: ck}), which gives:
\begin{equation}
    P_r(k)  = \left\{
    \begin{array}{ll}
        k(c_k - 2 c_{k+1} + c_{k+2}) &, \forall k \in \{1,...,L-2\} \\
        (L-1)(c_{L-1} - c_L) &, k=L-1 \\
        c_L &, k=L
    \end{array}
\right.
\end{equation}
This can be generalized as $P_r(k) = k \langle n_k \rangle /L$ for $k \in [1,L]$ using Eq.~\eqref{eq: statistical weights correlations}.
We define the probability $P_r(0)$ that the site does not belong to any cluster giving the property $\sum\limits_{k=0}^L P_r(k) = 1$.
Hence, the conditional probability, $Q(k)$ (Eq.~\eqref{eq:qk}), that a site belongs to a cluster of size $k$ knowing that this site is part of a cluster is:
\begin{equation}
    Q(k) = \frac{P_r(k)}{\sum\limits_{i=1}^L P_r(i)} = \frac{k \langle n_k \rangle}{\sum\limits_{i=1}^L i \langle n_i \rangle} = \frac{k \langle n_k \rangle}{\langle N \rangle} \,.
\end{equation}

\section{\label{app:dwandnk}Figure of mean number of domain walls and mean number of clusters}
\begin{figure}[H]
\includegraphics[width=0.5\textwidth]{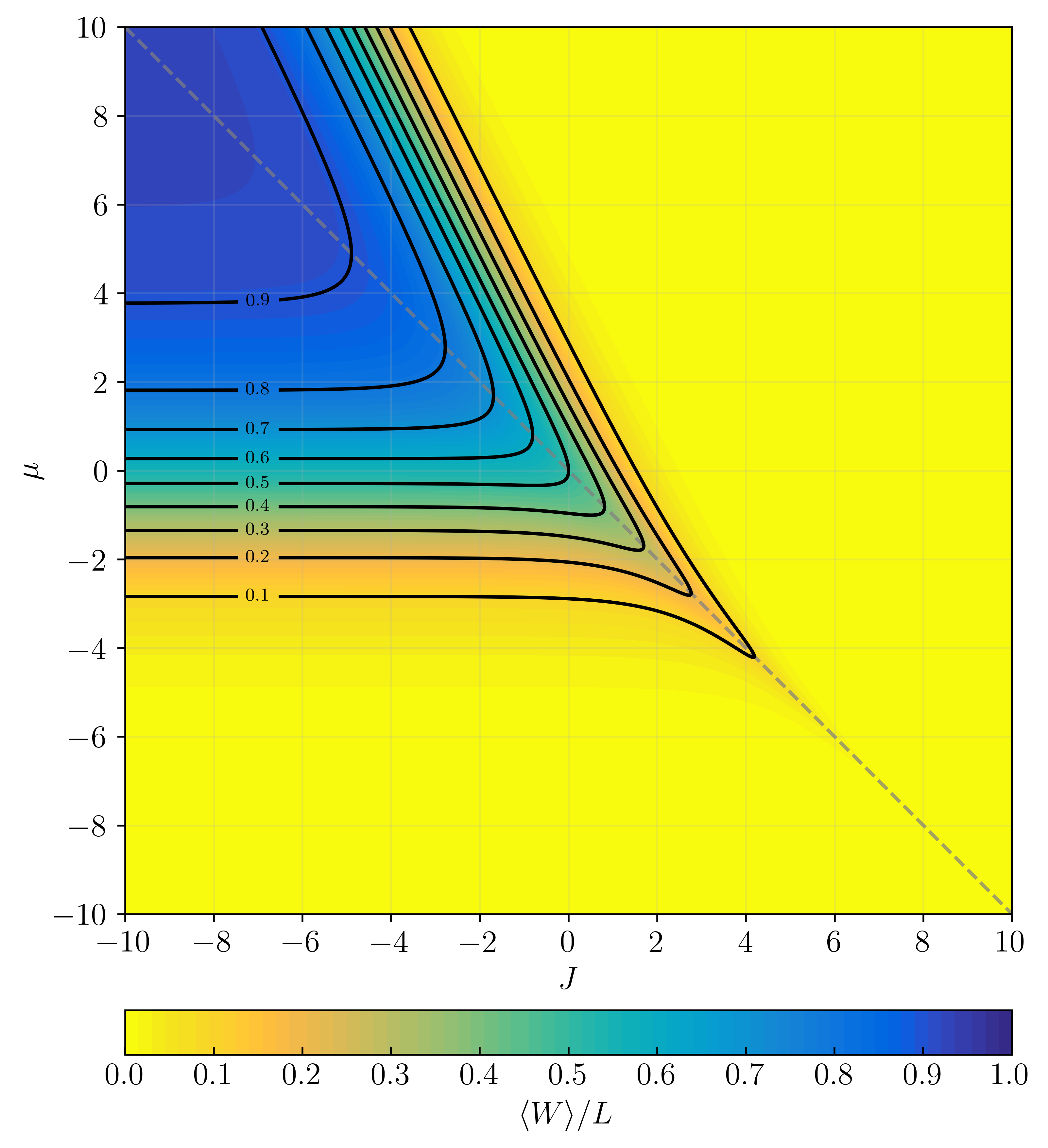}
\caption{Contour plot and heat map of the normalized mean number of domain walls at equilibrium, $\langle W \rangle /L$, as a function of the dimensionless interaction potential $J$ and chemical potential $\mu$ ranging from $-10$ to $10$, according to Eq.~\eqref{eq: domain walls} for a lattice size $L=13$.
The dashed gray line represents the Half-Filling of the lattice, where $J+\mu=0$.}
\label{fig:map dw}
\end{figure}
\begin{figure}[H]
\includegraphics[width=0.5\textwidth]{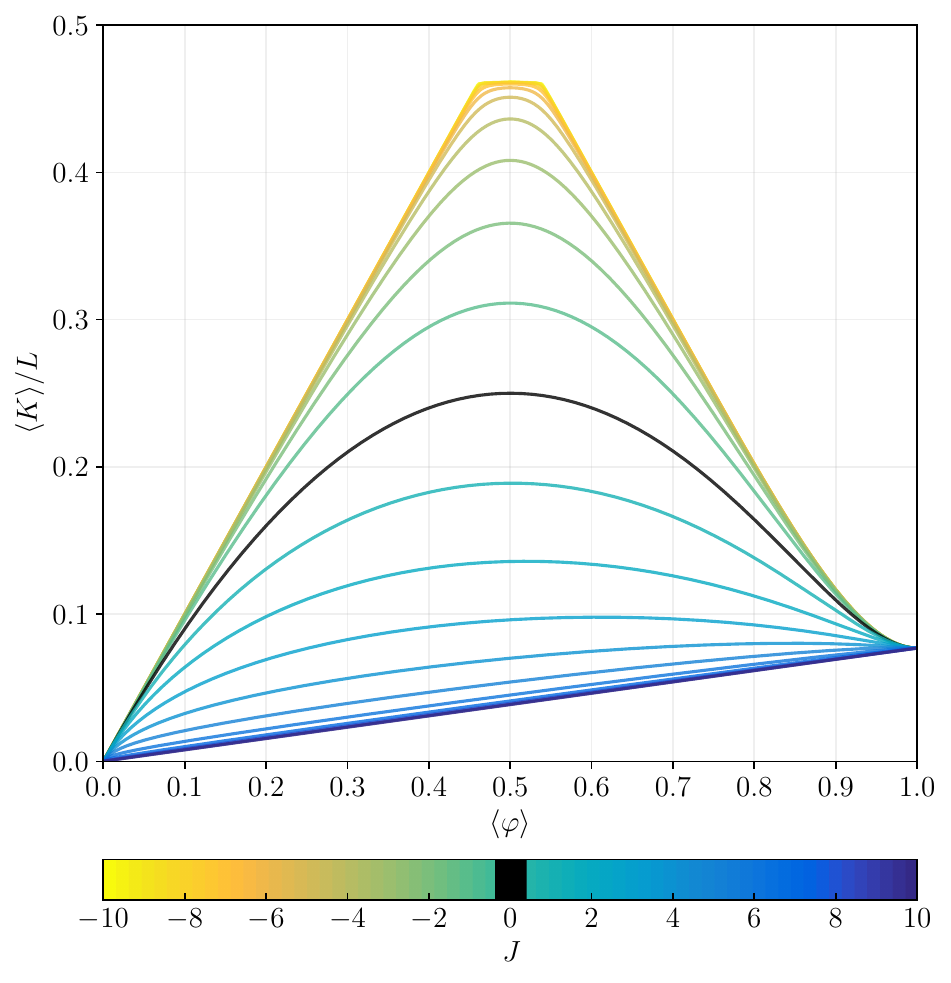}
\caption{Parametric plot of the normalized mean number of clusters at equilibrium, $\langle K \rangle /L$, as a function of the mean relative occupancy $\langle \varphi \rangle$, for a lattice of size $L=13$ according to the analytical expression in Eq.~\eqref{eq: number of cluster}.
For each curve,  the dimensionless interaction potential $J$ ranging from $-10$ (upmost curve) to $10$ (lowest curve) in steps of 1, is held fixed and $\mu$ is varied so that $\langle \varphi \rangle$ covers the interval $[0,1]$.
The black curve denotes the Hill-Langmuir case, $J=0$, computed from Eq.~\eqref{eq: mean number of clusters hl}.}
\label{fig: parametric number of cluster}
\end{figure}
\section{\label{sec: limiting cases}Limiting cases}
\subsection{Hill-Langmuir\label{sec: hill langmuir}}
The Hill-Langmuir (HL) regime represents an example of noniteracting particles, hence $J=0$. In this case $\lambda_+ = u_+= 1 + e^\mu$ and $\lambda_- = u_- = 0$, thus the partition function is simply $\Xi_{\text{HL}} = (1 + e^\mu)^L$.
In this case, the system is completely uncorrelated as $\xi_{\text{HL}} = 0$.
The mean relative occupancy can be derived from Eq.~\eqref{eq: mean relative occupancy}:
\begin{equation}
    \langle \varphi\rangle_\text{HL} = \frac{1}{1 + e^{-\mu}} \,.
    \label{eq: mean relative occupancy hl}
\end{equation}
Using this relation, we can now express the chemical potential as a function of the mean relative occupancy: 
\begin{equation}
    \mu = 2 \arctanh{(2\langle \varphi\rangle_{\text{HL}} -1)} \,.
    \label{eq: mu hl}
\end{equation}
We can now define all the previously established observables as a function of the mean relative occupancy.
Thus, the partition function becomes $\Xi_{\text{HL}} = (1 - \langle \varphi\rangle_\text{HL})^{-L}$.
We analytically continue the expression on the right-end-side of Eq.~\eqref{eq: domain walls} towards $J=0$, by taking the limit:
\begin{equation}
    \langle W \rangle_\text{HL} = 2L\langle\varphi\rangle_{\text{HL}} (1-\langle\varphi\rangle_{\text{HL}})  \,.
    \label{eq: domain walls hl}
\end{equation}
Using the same reasoning, the expression for the mean number of clusters is:
\begin{equation}
    \langle K \rangle_{\text{HL}} = L \langle \varphi \rangle_{\text{HL}}(1-\langle \varphi \rangle_{\text{HL}}) + \langle \varphi \rangle_{\text{HL}}^L \,.
    \label{eq: mean number of clusters hl}
\end{equation}
In the Hill-Langmuir regime, this two observables present interesting simple limits: the mean number of domain walls presents the same limit at the boundaries of occupancy since $\lim_{\langle \varphi \rangle_{\text{HL}} \to 0} \langle W \rangle_{\text{HL}} = \lim_{\langle \varphi \rangle_{\text{HL}} \to 1} \langle W \rangle_{\text{HL}} = 0 $,
whereas, $\lim_{\langle \varphi \rangle_{\text{HL}} \to 0} \langle K \rangle_{\text{HL}} = 0$, but $\lim_{\langle \varphi \rangle_{\text{HL}} \to 0} \langle K \rangle_{\text{HL}} = 1$, as the lattice configurations are reduced to a single system-sized cluster.

The expectation value of the cluster size without nearest neighbor interaction can be expressed using the mean number of clusters, Eq.~\eqref{eq: mean number of clusters hl}, as:
\begin{equation}
  \kappa_{\text{HL}} = \frac{L\langle \varphi\rangle_\text{HL}}{L \langle \varphi \rangle_{\text{HL}}(1-\langle \varphi \rangle_{\text{HL}}) + \langle \varphi \rangle_{\text{HL}}^L} \,.
  \label{eq: average cluster size hl}
\end{equation}
When examining the limiting cases, we obtain $\lim_{\langle \varphi \rangle_{\text{HL}} \to 0} \kappa_{\text{HL}} = 1 $ and $\lim_{\langle \varphi \rangle_{\text{HL}} \to 1} \kappa_{\text{HL}} = L$, respectively, as expected from Fig.~\ref{fig:parametric average cluster size}.

In the Hill-Langmuir regime, the correlation functions defined in Eq.~\eqref{eq: ck eigenvalues} can be expressed as a function of $\langle \varphi \rangle_{\text{HL}}$:
\begin{equation}
c_{k_\text{HL}} = \langle \varphi \rangle_\text{HL}^k \,.
\label{eq: correlation function hl}
\end{equation}
For this reason, in the Hill-Langmuir regime, it holds that $c_{k}^ \text{HL}\times c_{1}^\text{HL} = c_{k+1}^\text{HL}$ for $k\in\{1,\ldots,L\}$. 
Therefore, Eq.~\eqref{eq: statistical weights correlations} reduces to:
\begin{equation}
  \langle n_k \rangle_\text{HL} = \left\{
    \begin{array}{ll}
        L \langle \varphi \rangle_{\text{HL}}^k (1 - \langle \varphi \rangle_{\text{HL}})^{2} &, \forall k \in \{1,...,L-2\} \\
        L \langle \varphi \rangle_{\text{HL}}^{L-1} (1 - \langle \varphi \rangle_{\text{HL}}) &, k=L-1 \\
        \langle \varphi \rangle_{\text{HL}}^L &, k=L
    \end{array}
\right.
\label{eq: mean number of clusters of size k hl}
\end{equation}
We can remark that in the Hill-Langmuir regime, the mean number of clusters of size $k$ decreases exponentially as their size increases.

The cluster-size distribution then takes the form:
\begin{equation}
  P_\text{HL}(k) = \frac{\langle n_k \rangle_\text{HL}}{\langle K \rangle_\text{HL}} \,,
  \label{eq: probability occurrence cluster hl}
\end{equation}
and the site-weighted cluster size distribution is:
\begin{equation}
  Q_\text{HL}(k) = \frac{k\langle n_k \rangle_\text{HL}}{\langle N \rangle_\text{HL}} = \frac{k\langle n_k \rangle_\text{HL}}{L\langle \varphi \rangle_\text{HL}}\,.
  \label{eq: qk hl}
\end{equation}

The leading term in the asymptotic expansion for the CSD as $\langle\varphi\rangle_\text{HL}\to 0$ characterizes the cluster statistics in the low-occupancy regime:
\begin{equation}
P_\text{HL}(k)\sim 
\begin{cases}
\langle\varphi\rangle_\text{HL}^{k-1}-\langle\varphi\rangle_\text{HL}^k, & k \in \{1,\ldots,L-2\} \\
\langle\varphi\rangle_\text{HL}^{L-2}, & k = L-1 \\
\langle\varphi\rangle_\text{HL}^{L-1}/L, & k=L
\end{cases}
\label{eq: expension hl probability occurrence phi 0}
\end{equation}
This result indicates that, in the weak-occupancy limit, the statistics are dominated by single-particle clusters. 
However, for sufficiently large values of $L$, the asymptotic approximation also remains accurate at intermediate relative occupancy, provided that size-induced boundary effects can be neglected; in this intermediate regime, clusters of increasing size begin to emerge and compete with non-negligible weight, while remaining subdominant.
On the other hand, since the boundary at $k=L$ forces the distribution to collapse sharply, asymptotic series have limited validity in capturing the approach to the high-occupancy limit. 
A second-order asymptotic expansion provides a more accurate description of the CSD near this limit, $\langle \varphi\rangle_\text{HL}\to 1$, refining the leading-order result:
\begin{widetext}
\begin{equation}
P_\text{HL}(k)\sim 
\left\{
\begin{array}{ll}
L\left(1 -\langle\varphi\rangle_\text{HL}\right)^2, & k \in \{1,\ldots,L-2\} \\
L\left(\langle\varphi\rangle_\text{HL}-1\right)\left[L-2+(1-L)\langle\varphi\rangle_\text{HL}\right], & k = L-1  \\
1 -L\langle\varphi\rangle_\text{HL}\left(\langle\varphi\rangle_\text{HL}-1\right), & k=L
\end{array}
\right.
\label{eq: expansion hl probability occurrence phi 1}
\end{equation}
\end{widetext}

Nevertheless, it is difficult to capture finite-size effects, as evidenced by the fact that high-occupancy asymptotics remains accurate only in the regime of extremely large occupancy. However, it does correctly capture the essential feature that, as the relative occupancy increases, the system ultimately condenses into a single cluster of size $L$.

When carrying out the low occupancy limits of both $P_\text{HL}(k)$ and $Q_\text{HL}(k)$, we obtain
\begin{equation}
  \lim_{\langle \varphi \rangle_{\text{HL}} \to 0} P_\text{HL}(k) = \lim_{\langle \varphi \rangle_{\text{HL}} \to 0} Q_\text{HL}(k) = \delta_{k,1} \,,
  \label{eq: lim hl probability occurrence phi 0}
\end{equation}
where $\delta_{m,n}$ is the Kronecker symbol. This result indicates that in the limit of very weak occupancy (almost no particles on the lattice), the only clusters that survive are those made up of at most one particle.
On the other hand in the high occupancy limits we find
\begin{equation}
  \lim_{\langle \varphi \rangle_{\text{HL}} \to 1} P_\text{HL}(k) = \lim_{\langle \varphi \rangle_{\text{HL}} \to 1} Q_\text{HL}(k) = \delta_{k,L} \,,
  \label{eq: lim hl probability occurrence phi 1}
\end{equation}
indicating that in the limit of very large occupancy, the sole cluster that is left is the $L$-sized cluster.
It should be noted that both these distributions exhibit similar behaviors near the boundaries of occupancy, which is expected.
The limits found for the mean number of cluster of size $k$ are similar, as $\lim_{\langle \varphi \rangle_{\text{HL}} \to 0}\langle n_k \rangle_{\text{HL}} = 0$, while $\lim_{\langle \varphi \rangle_{\text{HL}} \to 1}\langle n_k \rangle_{\text{HL}} = \delta_{k,L}$ which confirms that there is only one cluster per configuration at high occupancies, namely the system sized cluster.
\begin{figure}[h]
\includegraphics[width=0.5\textwidth]{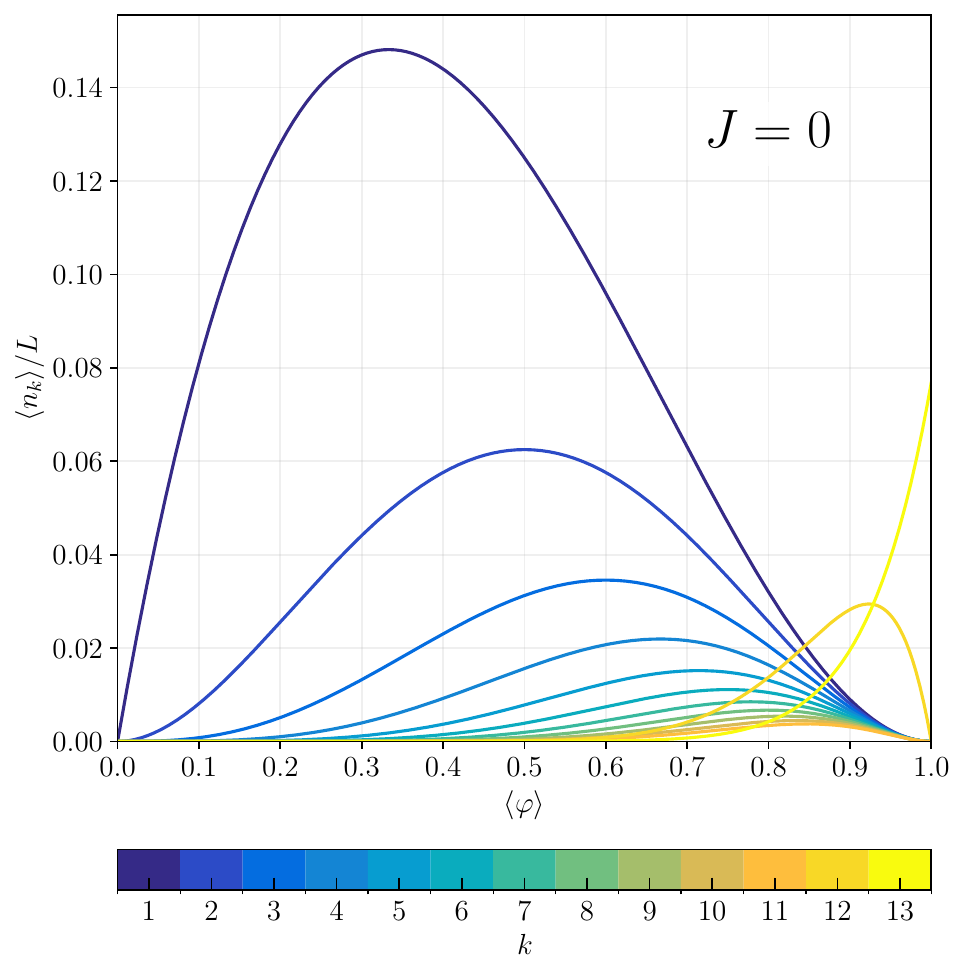}
\caption{Parametric plot of the normalized mean number of clusters of size $k$ at equilibrium, $\langle n_k \rangle /L$, as a function of the mean relative occupancy $\langle \varphi \rangle$, for a lattice of size $L=13$ in the Hill-Langmuir regime, $J=0$, according to the analytical expression in Eq.~\eqref{eq: mean number of clusters of size k hl}.
For each curve, the cluster size $k$ ranging from 1 (upmost curve) to $L=13$ (lowest curve) in steps of 1, is held fixed and $\mu$ is varied so that $\langle \varphi \rangle$ covers the interval $[0,1]$.}
\label{fig:parametric mean number of cluster of size J = 0}
\end{figure}
Fig.~\ref{fig:parametric mean number of cluster of size J = 0} shows the evolution in the Hill-Langmuir regime of the normalized number of clusters, $\langle n_k \rangle /L$, as a function of the relative occupancy for the cluster sizes, $k$, ranging from 1 to $L$. 
As the interaction potential is null, the number of clusters of a given size depends only on the occupancy and the system size. 
The number of clusters follows a skewed parabola, where the maxima shift toward larger occupancies as the cluster size increases.
As expected, clusters of size one are the most frequently found, especially before HF, and decrease for larger occupancy.
We can also remark that the curve for $k=2$ presents a maximum at HF and is symmetrical with respect to this maximum.
We observe that the curve corresponding to clusters of size \( k = L - 1 \) reaches its maximum at \( 1 - 1/L \), which is precisely the point where it intersects the curve associated with clusters of size \( L \).
For occupancy in the range of $\langle \varphi \rangle \geqslant 1 - (1/L)$, where, on average, only one site is empty, clusters of size $L-1$ and $L$ are most highly represented as they are the only ways to arrange bound particles.
At full occupancy, the $k=L$ curve reaches a normalized maximum of $1/L$ as the only possible cluster is one of system size.
\newline
Finally, we can derive an exact solution of the configuration-averaged cluster size, where Eq.~\eqref{eq: mean cluster size exact} simplifies to
\begin{widetext}
\begin{equation}
\begin{split}
    & \langle C \rangle_{\text{HL}} = L \langle \varphi \rangle_{\text{HL}}^L - 2^{1-L} \sum_{k=1}^{\left\lfloor \frac{L}{2}\right\rfloor} \binom{L}{2k} (2 \langle \varphi \rangle_{\text{HL}} -1)^{2k-1} \\
    & \times \left( (2\langle \varphi \rangle_{\text{HL}} -1)^{-2k} - \frac{4 \langle \varphi \rangle_{\text{HL}}^2 k (\langle \varphi \rangle_{\text{HL}} -1) [4k(\langle \varphi \rangle_{\text{HL}} -1) - L(2 \langle \varphi \rangle_{\text{HL}} -1) ] _{3}F_{2}\left(1,1,1-k; 2,2; \frac{4\langle \varphi \rangle_{\text{HL}}(\langle \varphi \rangle_{\text{HL}}-1)}{(1-2\langle \varphi \rangle_{\text{HL}})^2}\right)}{(2 \langle \varphi \rangle_{\text{HL}} -1)^{2}}
    -1 \right),
\end{split}
\label{eq: mean cluster size hl}
\end{equation}
\end{widetext}
where $_{3}F_{2}$ is the hypergeometric function of order $(3,2)$.
Taking the occupancy limits of this observable gives the expected result $\lim_{\langle \varphi \rangle_{\text{HL}} \to 0} \langle C \rangle_{\text{HL}} = 0$ as the system is empty and, $\lim_{\langle \varphi \rangle_{\text{HL}} \to 1} \langle C \rangle_{\text{HL}} = L$ as the system is full, containing $L$ particles forming a single cluster.

\subsection{Half-Filling\label{sec: half filling}}
As $J + \mu = 0$, Eq.~\ref{eq: mean relative occupancy} immediately yields $\langle \varphi \rangle_{\text{HF}} = 1/2$, hence the title of this subsection, \textit{half-filling regime}. In this regime, the eigenvalues of the transfer matrix are $\lambda_\pm = 1 \pm e^{-J/2}$ and $u_\pm = \frac{1}{2}(1 \pm e^{J/2})^2$.
Thus, the partition function is $\Xi_\text{HF} = (1 + e^{-J/2})^L + (1 - e^{-J/2})^L$.
The correlation length can be written as $\xi_\text{HF} = 1 / \log(\coth J/4)$.
Which allows us to give exact expressions for the number of domain walls using Eq.~\eqref{eq: domain walls},
\begin{equation}
  \langle W \rangle_\text{HF} = \frac{L}{e^J - 1} \left(\tanh \left(\frac{L}{2 \xi_\text{HF}} \right) e^{J/2} -1\right) \,,
  \label{eq: domain walls hf}
\end{equation} 
and the mean number of cluster with Eq.~\eqref{eq: number of cluster} as:
\begin{equation}
  \langle K \rangle_\text{HF} = \frac{2\Xi_\text{HF}^{-1} + \langle W \rangle_\text{HF}}{2} \,.
\end{equation}
Then Eq.~\eqref{eq: kappa} immediately yields an expression for the average cluster size using the mean number of clusters:
\begin{equation}
  \kappa_\text{HF} = \frac{L}{2\langle K \rangle_\text{HF}} = \frac{L}{2\Xi_\text{HF}^{-1} + \langle W \rangle_\text{HF}} \,.
  \label{eq: average cluster size hf}
\end{equation}

Carrying out the limits for the last three observables, we show that they are parity dependent in the anticooperative limit,
\begin{equation}
  \lim_{J \to - \infty} \langle W \rangle_{\text{HF}} = \left\{
    \begin{array}{ll}
        L &, L \text{ even,} \\
        L-1 &, L \text{ odd}
    \end{array}
\right.
\label{eq: lim hf - inf domain walls}
\end{equation}
\begin{equation}
  \lim_{J \to - \infty} \langle K \rangle_{\text{HF}} = \left\{
    \begin{array}{ll}
        L/2 &, L \text{ even,} \\
        (L-1)/2 &, L \text{ odd}
    \end{array}
\right.
\label{eq: lim hf - inf mean number of clusters}
\end{equation}
\begin{equation}
  \lim_{J \to - \infty} \kappa_{\text{HF}} = \left\{
    \begin{array}{ll}
        1 &, L \text{ even,} \\
        L/(L-1) &, L \text{ odd}
    \end{array}
\right.
\label{eq: lim hf - inf alpha}
\end{equation}
but parity independent in the cooperative limit:
\begin{equation}
  \lim_{J \to + \infty} \langle W \rangle_{\text{HF}} = 0 \,,
  \label{eq: lim hf + inf domain walls}
\end{equation}
\begin{equation}
  \lim_{J \to + \infty} \langle K \rangle_{\text{HF}} = 1/2 \,,
  \label{eq: lim hf + inf mean number of clusters}
\end{equation}
\begin{equation}
  \lim_{J \to + \infty} \kappa_{\text{HF}} = L \,.
  \label{eq: lim hf + inf alpha}
\end{equation}
The configurational implications of the parity (in)dependence will be discussed in details for the limits of the probability of occurrence of clusters of size $k$, $P_\text{HF}(k)$.
The mean number of clusters of size $k$, $\langle n_k \rangle$, is defined using the correlation functions introduced in Eq.~\eqref{eq: ck} as:
\begin{equation}
  \langle n_k \rangle_\text{HF} = \left\{
    \begin{array}{ll}
        L e^{-J} (\lambda_+^{L-k-1} + \lambda_-^{L-k-1}) / 2 \Xi_\text{HF} &, \forall k \in \{1,...,L-2\} \\
        L e^{-J}/ \Xi_\text{HF} &, k=L-1 \\
        1/\Xi_\text{HF} &, k=L
    \end{array}
\right.
\label{eq: statistical weights correlations hf}
\end{equation}
Building on the approach from the previous section, using the mean number of clusters of size $k$ and the mean number of clusters, we can define the CSD, $P(k)$, and SCSD, $Q(k)$, as:
\begin{equation}
  P_\text{HF}(k) = \frac{\langle n_k \rangle_\text{HF}}{\langle K \rangle_\text{HF}} \,,
  \label{eq: probability occurrence cluster hf}
\end{equation}
and:
\begin{equation}
  Q_\text{HF}(k) = \frac{2 k \langle n_k \rangle_\text{HF}}{L} \,.
  \label{eq: qk hf}
\end{equation}
When carrying out the limits of $P_\text{HF}(k)$ and $Q_\text{HF}(k)$, we find:
\begin{equation}
  \lim_{J \to + \infty} P(k)_{\text{HF}}= \lim_{J \to + \infty} Q(k)_{\text{HF}} = \delta_{k,L} \,.
  \label{eq: lim hf + inf probability of occurrence of cluster of size k}
\end{equation}
This result indicates that only clusters made up of $L$ particles will be observed in that regime.
This is indeed expected given that cooperativity is maximal in that case, confirming the switch-like behavior.
On the other hand, in the anticooperative limit:
\begin{equation}
  \lim_{J \to - \infty} P_{\text{HF}}(k) = \left\{
    \begin{array}{ll}
        \delta_{k,1} &, L \text{ even,} \\
        \frac{L-2}{L-1} \delta_{k,1} + \frac{1}{L-1} \delta_{k,2} &, L \text{ odd}
    \end{array}
\right.
\label{eq: lim hf - inf probability of occurrence of cluster of size k}
\end{equation}
This result arises because when $L$ is even, the only possible configuration at HF for $J \to - \infty$ is $\bm{\varphi} = \{1, 0, 1, 0, . . . , 1, 0 \}$, as particles repel each other; in other words, we are in the particle-hole dimer regime or perfectly mixed system.
Hence, the result $P_{\text{HF}}(k) = \delta_{k,1}$.
\newline
When $L$ is odd, however, a strict half-filling of the lattice is impossible.
The system then opts with equal probability either for a configuration made of $(L-1)/2$ size 1 clusters or for a configuration made of $(L-3)/2$ size 1 clusters and a size 2 additional cluster. 
The probability of a size 1 cluster is then $P_{\text{HF}}(1) = ((L-1)/2+(L-3)/2)/((L-1)/2 + (L - 3)/2 + 1) = (L - 2)/(L - 1)$ while the probability of a size 2 cluster is $P_{\text{HF}}(1) = 1/((L-1)/2+(L-3)/2+1) = 1/(L-1)$, whence the result for odd systems presented in Eq.~\eqref{eq: lim hf - inf probability of occurrence of cluster of size k}.
\newline
Using a similar reasoning, we find the anticooperative limit of $Q_\text{HF}(k)$ also presents a parity dependence:
\begin{equation}
  \lim_{J \to - \infty} Q_{\text{HF}}(k) = \left\{
    \begin{array}{ll}
        \delta_{k,1} &, L \text{ even,} \\
        \left(1 - \frac{2}{L}\right) \delta_{k,1} + \frac{2}{L} \delta_{k,2} &, L \text{ odd}
    \end{array}
\right.
\label{eq: lim hf - inf q k}
\end{equation}

The limits for $\langle n_k \rangle_{\text{HF}}$ present the same parity discrepancy, as the strongly cooperative limit is:
\begin{equation}
  \lim_{J \to + \infty}  \langle n_k \rangle_{\text{HF}} = \delta_{k,L} /2 \,.
  \label{eq: lim hf + inf number of cluster of size k}
\end{equation}
To understand this result, let us recall the probability of having a full lattice as $P_\varphi(1)=e^{L(J+\mu)}/\Xi$ (from Eq.~\eqref{eq: number of cluster}), which at HF gives $P_\varphi(1)=1/2$.
Moreover, the probability of having an empty lattice is also described by a single microstate corresponding to a null Hamiltonian, which immediately yields $P_\varphi(0)=1/\Xi=1/2$.
This is sufficient to show that the strong cooperative limit in the HF regime presents two filling configurations, \textit{i.e.}, empty and full lattices, that appear with the same probability.
Hence, we have $\langle n_L \rangle \to 1/2$ and $\langle n_k \rangle \to 0$ for all other cluster sizes.

Looking at the limits in the strong anticooperative limit gives:
\begin{equation}
  \lim_{J \to - \infty} \langle n_k \rangle_{\text{HF}} = \left\{
    \begin{array}{ll}
        \frac{L}{2}\delta_{k,1} &, L \text{ even,} \\
        \left( \frac{L}{2} - 1 \right) \delta_{k,1} + \frac{1}{2} \delta_{k,2} &, L \text{ odd}
    \end{array}
\right.
\label{eq: lim hf - inf number of cluster of size k}
\end{equation}
In the case of an even system, we have a perfect mixing of particles and holes, which means that the number of size-1 clusters in this type of configuration is
$L/2$ and there are no clusters of size greater than one, hence $\langle n_k \rangle \to L \delta_{k,1} /2$ in that limit.

When $L$ is odd, however, because of frustration, two types of configuration are possible: $\{1, 0, 1, 0, . . . , 1, 0, 0\}$ or $\{1, 0, 1, 0, . . . , 1, 0, 1\}$.
We shall call them \textit{S} (for single) and \textit{D} (for dimer), respectively.
Configuration \textit{S} has $(L -1)/2$ size-1 clusters while \textit{D} has $(L - 3)/2$ size-1 clusters and a single size-2 cluster (because of periodic boundary conditions).

It is easy by direct calculation to prove that the probability that the configuration \textit{S} occurs is $1/(2L)$ in the limit $J \to  \infty$ and that the probability that \textit{D} occurs is exactly the same.
As \textit{S} and \textit{D} are each $L$ times degenerated by rotation, the probability that a configuration of the \textit{S} or the \textit{D} type occurs is $1/2$, thus precluding the occurrence of any other configuration in that limit.
\newline
The average number of size-1 clusters is then $\langle n_1 \rangle = \frac{1}{2} \times \frac{L-1}{2} + \frac{1}{2} \times \frac{L-3}{2} = \frac{L}{2}-1$, while the average number of size-2 clusters is $\langle n_2 \rangle = \frac{1}{2} \times 0 + \frac{1}{2} \times 1 = 1/2$, hence the result quoted in Eq.~\eqref{eq: lim hf - inf number of cluster of size k}. 
The expression for the configuration-averaged cluster size, $\langle C \rangle$, greatly simplifies at HF given that $X=0$. 
We can then obtain 
\begin{equation}
    \langle C \rangle_\text{HF} = \frac{L \left(1 + \sum\limits_{k=1}^{\lfloor \frac{L}{2} \rfloor} \binom{L}{2k} \frac{e^{-Jk}}{k} \right)}{\Xi_\text{HF}} \,,
\label{eq: mean cluster size hf}
\end{equation}
where the sum appearing in the numerator may be expressed in terms of a hypergeometric function.
This quantity presents some interesting limits in the Half-Filling regime.
In the highly cooperative limit $J \to \infty$, we have $\langle C \rangle_\text{HF} \to L/2 = L \langle \varphi \rangle_\text{HF}$, highlighting once more that in this limit the configuration-averaged cluster size is analogous to the number of bound particles.
On the other hand, in the strongly anticooperative limit, parity effects appear, giving a different limit depending on the parity of $L$:
\begin{equation}
  \lim_{J \to - \infty} \langle C \rangle_\text{HF} = \left\{
    \begin{array}{ll}
        1 &, L \text{ even,} \\
        L/(L-1) &, L \text{ odd}
    \end{array}
\right.
\label{eq: lim hf - inf mean cluster size}
\end{equation}
The result in the case of an even lattice emphasizes the perfect mixing of particles and holes that governs the system for all occupancies inferior or equal to Half-filling.
The frustration of an odd system at HF is responsible for this second limit and is clearly visible in the insert of Fig.~\ref{fig:parametric mean cluster size}, characterized by the intermediate regime appearing for $\langle \varphi \rangle \in [(L-1)/2,(L+1)/2]$.
\subsection{Thermodynamic limit\label{sec: thermodynamic limit}}
As the lattice size $L$ tends to infinity and when the particle density (or occupancy) $\langle \varphi \rangle$ is neither zero nor one, the thermodynamic limit is reached.
Given that $\lambda_+ > \lambda_-$, expressions for the partition functions, the correlations, and the mean occupancy simplify as
\begin{equation}
  \Xi \sim \lambda_+^L \,,
  \label{eq: lt partition function}
\end{equation}
\begin{equation}
  c_k \sim e^{k \mu + J (k-1)} u_+ \lambda_+^{k-1}, \quad k \ll L \,,
  \label{eq: lt correlation function}
\end{equation}
and 
\begin{equation}
  \langle \varphi \rangle \sim \frac{e^\mu u_+}{\lambda_+^2} \,.
  \label{eq: lt mean relative occupancy}
\end{equation}
Now, taking into account the fact that $0 <e^{J+\mu} /\lambda_+ < 1$, it is easy to show that the inverse of the most probable cluster size, $\kappa$, is given by
\begin{equation}
  \frac{1}{\kappa} \sim \frac{\lambda_+}{u_+(\lambda_+ - \lambda_-)} = 1 - \frac{e^{J+\mu}}{\lambda_+} \,,
  \label{eq: lt average cluster size}
\end{equation}
as $L \to \infty$.
\newline
Thanks to the results obtained above, we are in a position to obtain the cluster size statistics, that is, the probability $P(k)$ that a cluster has size $k$ in the thermodynamic limit.
We first calculate the average number of clusters with size $k$, $\langle n_k \rangle$. We find,
\begin{equation}
  \langle n_k \rangle \sim L \langle \varphi \rangle \left( \frac{e^{J+\mu}}{\lambda_+} \right)^{k-1} \left( 1 - \frac{e^{J+\mu}}{\lambda_+}\right)^2 \,.
  \label{eq: lt mean number of clusters of size k}
\end{equation}
Using Eq.~\eqref{eq: probability occurrence cluster} giving the probability $P(k)$ that a cluster has a size $k$ and Eq.~\eqref{eq: kappa} giving the most probable cluster size encountered in the system as the ratio of the mean number of bound particles to the mean number of clusters, and the result that, as $L \to \infty$, Eq.~\eqref{eq: lt average cluster size} holds, we eventually find:
\begin{equation}
  P(k) \sim \frac{1}{\kappa} \left(1 - \frac{1}{\kappa} \right)^{k-1} \,.
  \label{eq : lt probability cluster size k with average cluster size} 
\end{equation}
Where the most probable cluster size is given in the thermodynamic limit by Eq.~\eqref{eq: lt average cluster size}. 
This expression is correctly normalized $(k \in \mathbb{N}^*)$ and self-consistent with the property that $\kappa$ is the expectation value of the cluster size with respect to the CSD.
As we readily see, it is geometric, which means that in the thermodynamic limit, the probability to find clusters with size $k$ decreases exponentially as $k$ increases.
Moreover, we can show that $\kappa$ can be expressed solely as a function of the mean relative occupancy $\langle \varphi \rangle$ and the
interaction potential $J$ as
\begin{equation}
  \frac{1}{\kappa} \sim \frac{\sqrt{1 + 4 \langle \varphi \rangle(1-\langle \varphi \rangle)(e^J -1)} -1}{2\langle \varphi \rangle(e^J -1)} \,,
  \label{eq: lt average cluster size function of occupancy}
\end{equation}
as $L \to \infty$.
Expressions similar to Eqs.\eqref{eq : lt probability cluster size k with average cluster size} and \eqref{eq: lt average cluster size function of occupancy} have already been obtained in \cite{yilmaz2005exact} within the framework of the canonical ensemble.
This is indeed expected given the equivalence of the statistical ensembles in the thermodynamic limit.
Let us note further that, for $k > 1$, the probability $P(k)$ achieves a maximum for a value of the average cluster size precisely given by $\kappa^* = k$, as expected.
Upon using Eq.~\eqref{eq: lt average cluster size function of occupancy}, we see that this corresponds to a mean occupancy given by $\langle \varphi \rangle^*(k) = k(k-1)/(k^2 +e^J -1)$.
Therefore, in the thermodynamic limit $L \to \infty $, the probability to find a cluster of size $k$ is maximum when the mean occupancy is given by $\langle \varphi \rangle^*(k)$ where
\begin{equation}
  \langle \varphi \rangle^* (k) = \frac{k(k-1)}{k^2 + e^J - 1} \,,
  \label{eq: lt mean relative occupancy function of k}
\end{equation}
and this $k$-dependent maximum is given by
\begin{equation}
  P^*(k) = \frac{1}{k} \left(1 - \frac{1}{k} \right)^{k-1} \sim \frac{e^{-1}}{(k-1)} \,,
  \label{eq: lt probability cluster size k function of k}
\end{equation}
where the last expression is valid for large values of $k$.
\begin{figure*}[t]
\includegraphics[width=1\textwidth]{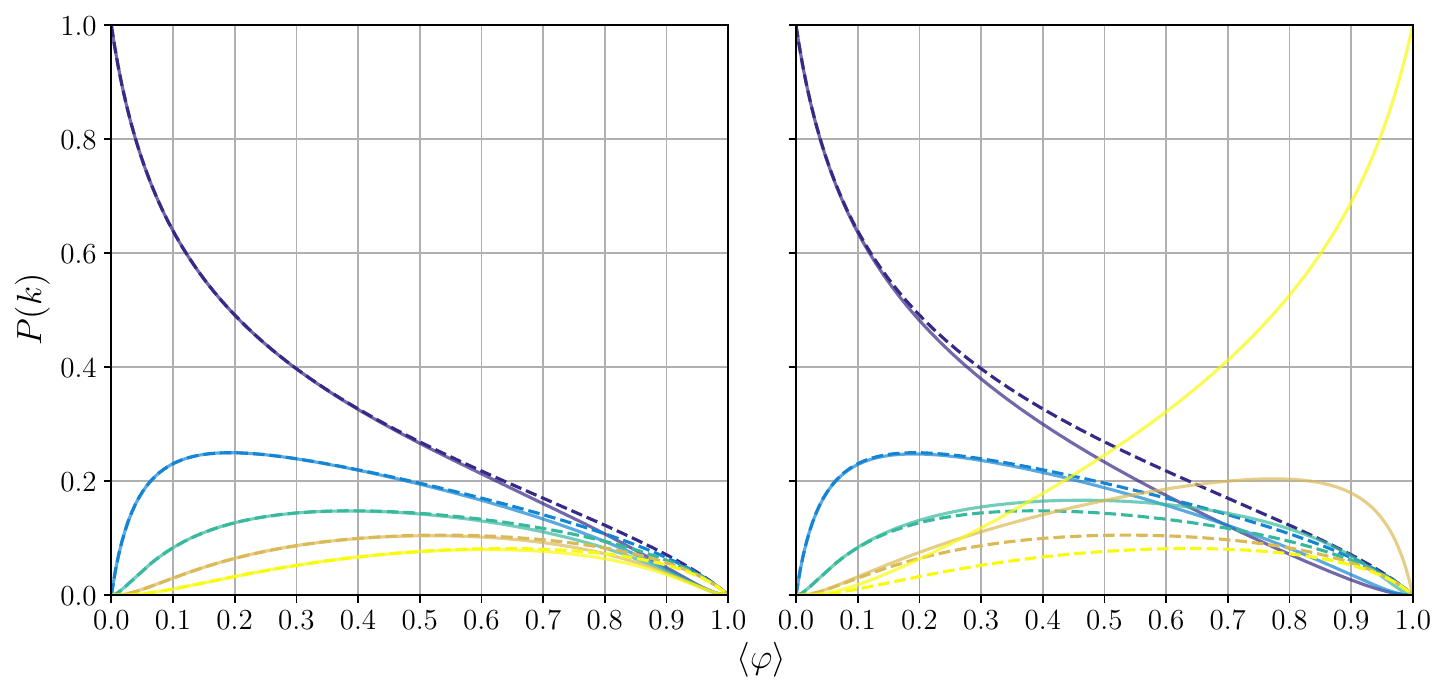}
\caption{Evolution of the equilibrium probability of occurrence of a cluster of size $k$, $P(k)$ as a function of the mean relative occupancy, $\langle \varphi \rangle$, calculated in the general case with Eq.~\eqref{eq: probability occurrence cluster} (solid lines) and in the thermodynamic limit with Eqs.~\eqref{eq : lt probability cluster size k with average cluster size} and \eqref{eq: lt average cluster size function of occupancy} (dashed lines), for a dimensionless interaction potential $J=2$, and different values of cluster size $k$ increasing from 1 to 5 (purple, blue, green, brown, yellow).
The left panel represents a lattice size of $L=13$ and the right panel a size of $L=5$.}
\label{fig:lt-probability-of-occurence-of-cluster-of-size-k}
\end{figure*}
As can be seen from Fig.~\ref{fig:lt-probability-of-occurence-of-cluster-of-size-k}, the thermodynamic limit ($L \to \infty$) works already well for $L = 13$ and for not too high values of the mean relative occupancy $\langle \varphi \rangle $, but fails to reproduce the correct trend as $\langle \varphi \rangle \to 1$. 
For $L = 5$, on the contrary, there is clearly no agreement between the thermodynamic limit results and the exact ones.
Note that in the Hill-Langmuir regime ($J \to 0$), Eq.~\eqref{eq: lt average cluster size function of occupancy} yields $\kappa^{-1} = 1 - \langle \varphi \rangle$.
In that limit, the probability that a cluster has size $k$ then becomes $P(k) \sim (1 - \langle \varphi \rangle) \langle \varphi \rangle^{k-1}$, an expression that can also be obtained by taking the limit ($L \to \infty$) in the exact Hill-Langmuir formula \ref{eq: probability occurrence cluster hl}.
Finally, the correction term used to derive the general equation for the mean number of clusters, $\langle K \rangle$, (Eq.~\eqref{eq: number of cluster}) is null, as we are in the thermodynamic limit, thus $P_\varphi(1) \sim 0$ when $L \to \infty$.
More generally, only configurations with a number of bound particles that scales with $L$  survive with non-zero probability in the thermodynamic limit.
Hence, the mean number of domain walls in the thermodynamic limit is:
\begin{equation}
  \langle W \rangle \sim 2 \langle K \rangle = \frac{2 \kappa}{\langle N \rangle} \,.
  \label{eq: lt domain walls}
\end{equation}
\section{\label{app:hfinhl}Half-Filling in the Hill-Langmuir regime}
The Hill-Langmuir regime at Half-Filling is the simplest regime, characterized by the equality $J + \mu = 0$ and the property $\langle \varphi \rangle = 1/2$.
In that case, all configurations $\bm{\varphi}$ have the same probability to occur at thermal equilibrium.
A site then has the same probability of being occupied or empty, resulting in half-filling of the lattice.
Thus, exact results can be derived for all observables previously described using clusters and their size.
As $\lambda_- = 0$ in the HL regime, the partition function depends only on $\lambda_+ = 2$, giving $\Xi = 2^L$.
The mean number of domain walls given by Eq.~\eqref{eq: domain walls} yields:
\begin{equation}
  \langle W \rangle = \frac{L}{2}\,.
  \label{eq: domain walls hl hf}
\end{equation}
This allows us to write the mean number of clusters as:
\begin{equation}
  \langle K \rangle = \frac{L}{4} + 2^{-L} \underset{L \gg 1}{\sim} \frac{L}{4} \,.
  \label{eq: number of clusters hl hf}
\end{equation}
The correlation functions are easily obtained using Eq.~\eqref{eq: correlation function hl} as:
\begin{equation}
  c_k = 2^{-k}\,.
\label{eq: correlation function hl hf}
\end{equation}
Using Eq.~\eqref{eq: mean number of clusters of size k hl} we obtain the mean number of clusters of size $k$ as:
\begin{equation}
  \langle n_k \rangle = \left\{
    \begin{array}{ll}
        L /2^{k+2}  &, \forall k \in \{1,...,L-2\} \\
        L/2^L &, k=L-1 \\
        1/2^L &, k=L
    \end{array}
\right.
\label{eq: statistical weights correlations hl hf}
\end{equation}
In the HL regime at HF, the mean number of clusters of size $k$ times the partition function,$\langle n_k \rangle \times \Xi$, gives the exact number of clusters of size $k$ in a system of size $L$, given that the Boltzmann weight $e^{-\beta\mathcal{H}(\bm{\varphi})} = 1$ for $J = \mu = 0$.
This number of clusters of size $k$ decreases exponentially when the size $k$ of the cluster increases.
Using the mean number of clusters of size $k$,$\langle n_k \rangle$, and the mean number of clusters, $\langle K \rangle$, we obtain the CSD in the Hill-Langmuir regime at HF:
\begin{equation}
  P(k) = \left\{
    \begin{array}{ll}
        \frac{L 2^{L-k-2} }{L 2^{L-2} +1}  &, \forall k \in \{1,...,L-2\} \\
        \frac{L}{L 2^{L-2} +1} &, k=L-1 \\
        \frac{1}{L 2^{L-2} +1} &, k=L
    \end{array}
\right.
\label{eq: probability of occurrence cluster k hl hf}
\end{equation}
For a large enough lattice size $L$, the behavior of this probability is exponentially decreasing with increasing cluster size:
\begin{equation}
  P(k) = \left\{
    \begin{array}{ll}
        2^{-k}  &, \forall k \in \{1,...,L-2\} \\
        2 \times 2^{1-L} &, k=L-1 \\
        4 \times 2^{-L} / L &, k=L
    \end{array}
\right.
\label{eq: probability of occurrence cluster k hl hf large L}
\end{equation}
Using the definition of the SCSD at HF in Eq.~\eqref{eq: qk hf} and the mean number of cluster of size $k$, $\langle n_k \rangle$ given by Eq.~\eqref{eq: statistical weights correlations hl hf}, the SCSD is simply
\begin{equation}
  Q(k) = \left\{
    \begin{array}{ll}
        k / 2^{k+1}  &, \forall k \in \{1,...,L-2\} \\
        (L-1) / 2^{L-1} &, k=L-1 \\
        1/2^{L-1} &, k=L
    \end{array}
\right.
\label{eq:q k hl hf}
\end{equation}
where this distribution follows an exponential decay multiplied by a linear increase.
Using Eq.~\eqref{eq: average cluster size correlation}, we can use the correlation functions to determine the most probable cluster size, which can be written as:
\begin{equation}
  \kappa = \frac{L 2^{L-1}}{L 2^{L-2} +1} \underset{L \gg 1}{\sim} 2 \,,
\label{eq: kappa hl hf}
\end{equation}
where we have used the exact results provided in Eq.~\eqref{eq: probability of occurrence cluster k hl hf} to obtain the first result, which simplifies to $2$ when $L$ is large enough.
Finally, using Eq.~\eqref{eq: mean cluster size exact}, the configuration-averaged cluster size becomes
\begin{equation}
  \langle C \rangle = \frac{L \left( 1 + \sum\limits_{k+1}^{\left\lfloor \frac{L}{2} \right\rfloor} \frac{1}{k} \binom{L}{2k} \right) }{2^L} \underset{L \gg 1}{\sim} \frac{L + 2^{L+1}}{2^{L}} \sim 2 ,
\label{eq: averaged cluster size hl hf}
\end{equation}
interestingly the mean cluster size, $\kappa$, and the configuration-averaged cluster size, $\langle C \rangle$, present the same exact limits in the large lattice limit.
In this regime, the mean number of particles is $\langle N \rangle = \langle W \rangle = L/2$, which means that each particle shares an edge with an empty site. 
For a sufficiently large lattice, the mean number of clusters is $\langle K \rangle = \langle N \rangle /2 = L/4$, and with $\langle C \rangle = \kappa = 2$, we find that in this limit, on average, the system is composed of $L/4$ clusters of size $2$ which are all separated by two empty sites at each boundary.
\section{\label{app:tableobs}Table of observables}

\begin{table}[H]
    \centering
    \begin{tabular}{| c | c |}
        \hline
        Observable & Definition\\
        \hline \hline
        $\langle \varphi \rangle$ & Mean relative occupancy\\
        \hline
        $\langle N \rangle$ & Mean number of bound particles\\
        \hline
        $\langle W \rangle$ & Mean number of domain walls\\
        \hline
        $\langle C \rangle$ & Configuration-averaged cluster size\\
        \hline
        $\langle K \rangle$ & Mean number of cluster\\
        \hline
        $P(k)$ & Cluster-site distribution (CSD)\\
        \hline
        $Q(k)$ & Site-weighted cluster-site distribution (SCSD)\\
        \hline
        $\kappa$ & Most probable cluster size\\
        \hline
        $\langle n_k \rangle$ & Mean number of clusters of size $k$\\
        \hline
        $c_k$ & $k$-site cluster correlation functions\\
        \hline
        $\Xi$ & Partition function\\
        \hline
    \end{tabular}
    \caption{Summary of observables and quantities with their definitions.}
    \label{tab:summaryobservables}
\end{table}

\bibliography{bib_cluster}

\end{document}